\documentclass[10pt]{article}
\usepackage[square,authoryear]{natbib}
\usepackage{graphicx}
\usepackage{stackrel}
\usepackage{cancel}
\usepackage[all]{xy}
\usepackage{epstopdf,framed}
\usepackage{pdfsync}
\usepackage{amscd}
\usepackage{mathrsfs}
\RequirePackage{color}
\usepackage{amsmath,bm}
\usepackage{amssymb}
\usepackage{amsfonts}
\usepackage{url}
\newtheorem{theorem}{Theorem}[section]

\newtheorem{definition}[theorem]{Definition}

\newcommand{\bfi}{\bfseries\itshape}
\newtheorem{remark}[theorem]{Remark}

\def\intprod{\mathbin{\hbox to 6pt{ \vrule height0.4pt width5pt depth0pt
\kern-.4pt \vrule height6pt width0.4pt depth0pt\hss}}}
\begin{document}

\title{Dirac structures in nonequilibrium thermodynamics for simple open systems}
\vspace{-0.2in}

\date{July 15, 2019}

\author{\hspace{-1cm}
\begin{tabular}{cc}
Fran\c{c}ois Gay-Balmaz &
Hiroaki Yoshimura
\\  CNRS, LMD, IPSL & School of Science and Engineering
\\ Ecole Normale Sup\'erieure & Waseda University
\\  24 Rue Lhomond 75005 Paris, France & Okubo, Shinjuku, Tokyo 169-8555, Japan \\ gaybalma@lmd.ens.fr & yoshimura@waseda.jp\\
\end{tabular}\\\\
}

\maketitle
\vspace{-0.3in}

\begin{center}
\abstract{Dirac structures are geometric objects that generalize Poisson structures and
presymplectic structures on manifolds. They naturally appear in the formulation of
constrained mechanical systems and play an essential role in structuring a dynamical system through the energy flow between its subsystems and elements. In this paper, we show that the evolution equations for open thermodynamic systems, i.e., systems exchanging heat and matter with the exterior, admit an intrinsic formulation in terms of Dirac structures. We focus on simple systems, in which the thermodynamic state is described by a single entropy variable. A main difficulty compared to the case of closed systems lies in the explicit time dependence of the constraint associated to the entropy production. We overcome this issue by working with the geometric setting of time-dependent nonholonomic mechanics. We define three type of Dirac dynamical systems for the nonequilibrium thermodynamics of open systems, based either on the generalized energy, the Lagrangian, or the Hamiltonian. The variational formulations associated to the Dirac systems formulations are also presented.}
\end{center}


\section{Introduction}\label{Sec_1}

Nonequilibrium thermodynamics is a phenomenological theory that aims to identify and
describe the relations among the observed macroscopic properties of a physical system and to
determine the macroscopic dynamics of this system with the help of fundamental laws (e.g. \cite{StSc1974}). The field of nonequilibrium thermodynamics naturally includes macroscopic disciplines such as classical mechanics, fluid dynamics, elasticity, and
electromagnetism. The main feature of nonequilibrium thermodynamics is the occurrence of the various irreversible processes, such as friction, mass transfer, and chemical reactions, which are accompanied by entropy production. In particular, a crucial class of nonequilibrium systems in many applications is no doubt that of open systems, in which there is an exchange of energy and matter between the system and the surroundings through its ports.

The classical theory of nonequilibrium thermodynamics emerged with the work of \cite{Onsager1931} on the reciprocal relations connecting the coefficients, which appear in the linear phenomenological relations between irreversible fluxes and thermodynamic forces. We refer to the classical books \cite{deGrootMazur1969, GlPr1971, StSc1974, Woods1975, Lavenda1978, KoPr1998}, for the general setting and various applications of nonequilibrium thermodynamics.

Several \textit{variational formulations} have been developed in the context of nonequilibrium thermodynamics, such as those based on the \textit{principle of least dissipation of energy} (\cite{Onsager1931,OnMa1953,MaOn1953}) and those based on the \textit{principle of minimum entropy production} (\cite{Prigogine1947,GlPr1971}). The former utilized the reciprocity appeared in the linear phenomenological relations, while the latter was generalized in \cite{Ziegler1968} to the case of systems with nonlinear phenomenological relations. We refer to \cite{Gyarmati1970, Lavenda1978} for more information on the variational formulations related to nonequilibrium thermodynamics.

On the other hand, the \textit{geometric formulation} of \textit{equilibrium} thermodynamics was mainly studied in the context of contact geometry, see \cite{He1973,Mr1978,Mr1980,MrNuScSa1991}, following the initial works of \cite{Gibbs1873a,Gibbs1873b,Ca1909}. In this geometric setting, thermodynamic properties are encoded by Legendre submanifolds of the thermodynamic phase space. A step toward a geometric formulation of irreversible processes was made in \cite{EbMaVa2007} by
lifting port-Hamiltonian systems to the thermodynamic phase space. The underlying geometric
structure in this construction is again a contact form.

\paragraph{Variational formulation and Dirac structures in nonequilibrium thermodynamics.}
A novel variational formulation for the nonequilibrium thermodynamics of finite dimensional and continuum closed systems has been proposed in \cite{GBYo2017a,GBYo2017b}. This variational formulation is an extension of the celebrated Hamilton principle of classical mechanics, which includes the irreversible processes and the entropy production equation. Namely, this formulation is based on a type of the Lagrange-d'Alembert principle associated with a special class of nonlinear constraints, which is called \textit{nonlinear constraints of thermodynamic type}. This type of variational formulation involves two kinds of constraints, namely, the one imposed on the critical curve and the other on the variations. In the context of nonequilibrium thermodynamics, thanks to the concept of thermodynamic displacement introduced in \cite{GBYo2017a,GBYo2017b} these two constraints are related in a systematic way as
\begin{equation}\label{relation_CV_CK}
\sum_\alpha J_\alpha\dot\Lambda^\alpha\;\; \leadsto\;\; \sum_\alpha J_\alpha\delta\Lambda^\alpha,
\end{equation}
with $J_\alpha$ the thermodynamic flux and $\Lambda^\alpha$ the thermodynamic displacement of the process $\alpha$.

 As shown in \cite{GBYo2018c} to this \textit{variational formulation} is naturally associated a \textit{geometric structure} given by a Dirac structure that allows to systematically formulate the equations for {\it simple isolated} thermodynamic systems as Dirac dynamical systems. In absence of irreversible process, this geometric formulation consistently recovers the standard Hamiltonian formulation of classical mechanics.

Let us recall that Dirac structures are geometric objects that extend the notion of presymplectic and Poisson structures and were introduced in \cite{Cour1990,CoWe1988}. They have been successfully used for the formulation of electric circuits, \cite{vdSMa1995a}; \cite{BlCr1997}, nonholonomic mechanical systems, \cite{vdSMa1995b}, as well as constrained implicit Lagrangian systems \cite{YoMa2006a,YoMa2006b}.

The variational formulation of nonequilibrium thermodynamics mentioned above has been extended to the case of discrete (finite dimensional)  \textit{open systems} in \cite{GBYo2018a}, see also \cite{GBYo2019} for a review of both closed and open systems. For such open systems, the exchange of matter between the system and the exterior through it ports induces fundamental changes in the structure of the constraints, since they become explicitly dependent on time and the relation \eqref{relation_CV_CK} is modified by additional terms that depend on the exterior of the system. The variational setting developed in \cite{GBYo2018a} includes these main features.

It is natural to ask if the evolution equations for thermodynamic system can be formulated in the context of Dirac structures, as in nonholonomic mechanics. For the case of simple and adiabatically closed systems, the present authors have shown in \cite{GBYo2018c} the construction of Dirac structures in nonequilibrium thermodynamics together with the associated Dirac dynamical systems. 

In this paper, we will show an extension to the case of simple open systems. In order to carry out this, we will use the geometric setting of time-dependent nonholonomic mechanics, which is based on the \textit{extended configuration manifold} $Y:=\mathbb{R}\times \mathcal{Q}\ni (t,x)$,  seen as a trivial vector bundle $Y=\mathbb{R}\times\mathcal{Q}\rightarrow \mathbb{R}$, $(t,x)\mapsto t$, over $\mathbb{R}$. This is the geometric setting of classical field theories as it applies to time-dependent mechanics (see, \cite{GIMM1997}). Even though the bundle is trivial and the base is one-dimensional, having this abstract point of view turns out to be crucial for our developments, since it directly provides us with the intrinsic concept of the covariant energy, covariant Hamiltonian, as well as the covariant Pontryagin bundle, associated with $Y=\mathbb{R}\times\mathcal{Q}\rightarrow \mathbb{R}$, which turn out to be fundamentals of our Dirac formulation.

\paragraph{Organization of the paper.} We start by quickly recalling the notion of Dirac structures and Dirac dynamical systems in nonholonomic mechanics, as well as the associated variational formulations, for the case in which the kinematic constraints are linear with respect to velocities. Then, we briefly indicate some of the difficulties that need to be overcome in the geometric formulation, when passing from adiabatically closed to open systems. In \S\ref{Sect_FundSet} we first make a short review on the variational formulation for simple open systems. Then we describe an abstract variational setting that reflects the main feature of our variational formulation and explain how it applies to open systems. In \S\ref{Section_3}, we describe an abstract setting involving two geometric objects, $C_V$ and $C_K$, associated to the constraints used in  the variational formulation of open systems and clarify geometrically the link between them in thermodynamics. Given a variational constraint $C_V$,  we construct a distribution on the covariant Pontryagin bundle, from which a Dirac structure is obtained. We then formulate the Dirac dynamical system associated to this Dirac structure and to the covariant generalized energy, and develop the evolution equations governed by this system. Two other Dirac formulations are also presented, which are respectively associated to the Lagrangian and the Hamiltonian, together with their associated variational formulations of Lagrange-d'Alembert-Pontryagin type. From these developments, we deduce in \S\ref{Section_4} the Dirac system formulation for open thermodynamic systems, and explicitly show that the complete set of evolution equations for the system can be obtained from the Dirac formulation. We also interpret physically the variables involved in the Dirac system formulation.

\paragraph{Dirac structures, Dirac dynamical systems, and application in nonholonomic mechanics.} Let $M$ be a manifold and consider the Pontryagin vector bundle $TM \oplus T^*M$ over $M$ endowed with the symmetric fiberwise bilinear form
\[
\left\langle \!\left\langle ( u_m, \alpha _m), (v_m, \beta _m) \right\rangle \! \right\rangle = \left\langle\beta _m, u_m\right\rangle + \left\langle \alpha _m, v_m \right\rangle ,
\]
for all $( u_m, \alpha _m), (v_m, \beta _m) \in T_mM\oplus T^*_mM$.
A \textit{Dirac structure on  $M$} is by definition a vector subbundle $D \subset TM \oplus T^*M$, such that $D^\perp=D$ relative to $\left\langle \!\left\langle \cdot ,\cdot \right\rangle \! \right\rangle $, see \cite{Cour1990}.

For example, given a two-form $ \omega \in \Omega ^2 (M)$ and a distribution $ \Delta_M $ on $M$ (i.e., a vector subbundle of $TM$), the subbundle $D_{ \Delta _M} \subset TM \oplus T^*M$ defined by, for each $m \in M$,
\begin{equation}\label{def_Dirac}
\begin{aligned}
D_{ \Delta _M}(m) :
&=\big\{ (v_{m}, \alpha_{m}) \in T_{m}M  \times T^{\ast}_{m}M  \; \mid \;  v_{m} \in \Delta_M(m) \; \text{and} \\ 
& \qquad \qquad \qquad \left\langle \alpha_{m},w_{m} \right\rangle =\omega(m)(v_{m},w_{m}) \; \;
\mbox{for all} \; \; w_{m} \in  \Delta _M(m) \big\}
\end{aligned}
\end{equation}
is a Dirac structure on $M$. In this paper, we shall extensively use this construction of Dirac structure. 

\begin{remark}[Integrability]{\rm A Dirac structure is called \textit{integrable} if it is involutive with respect to the \textit{Dorfman bracket} defined on the space of sections of $TM\oplus T^*M$ by
\[
[(X_1,\alpha_1),(X_2,\alpha_2)]:= \big([X_1,X_2], \pounds_{X_1}\alpha_2-\pounds_{X_2}\alpha_1 + \mathbf{d}\langle \alpha_1, X_2\rangle \big).
\]
Note that this bracket is not skew-symmetric but satisfies a Jacobi identity. We refer to \cite{KS2013} for references and historical review on this bracket. What we call Dirac structures are sometime referred to as {\it almost} Dirac structures in the literature, where  the name Dirac structure is only used for the integrable case. Since we are not concerned with the integrability condition in our applications to thermodynamics, we utilize for shortness the terminology 'Dirac structures' even for the nonintegrable cases, following \cite{YoMa2006a}.}
\end{remark}

Given a Dirac structure $D$ on $M$ and a function $ F: M \rightarrow \mathbb{R}$, the associated \textit{Dirac dynamical system} for a curve $m(t) \in M$ is
\begin{equation}\label{DDS}
\big(\dot m(t), \mathbf{d} F(m(t)) \big) \in D(m(t)),
\end{equation} 
where $\dot m(t)=\frac{d}{dt}m(t)$.
Note that in general \eqref{DDS} is a system of implicit differential-algebraic equations and the questions of existence, uniqueness or extension of solutions for a given initial condition can present several difficulties.

The formulation \eqref{DDS} presents a unified treatment of formulating the equations of nonholonomic systems with a (possibly degenerate) Lagrangian. Let us quickly recall how this proceeds in the following. Consider a mechanical system with configuration manifold $Q$, a Lagrangian $L:TQ \rightarrow \mathbb{R}  $ and a constraint distribution $ \Delta_Q \subset TQ$. Let $P:= TQ \oplus T^*Q$ be the Pontryagin bundle of $Q$ and define the induced distribution $ \Delta _P(q,v,p):= \left( T_{(q,v,p)} \pi _{(P, Q)} \right) ^{-1} ( \Delta _Q(q))$, where $ \pi _{(P,Q)}: P \rightarrow Q$ is the projection defined by $ \pi _{(P,Q)}(q,v,p)=q$, where we use the local coordinates $q$, $(q,v)$ and $(q,p)$, respectively,  for $Q$, the tangent bundle $TQ$, and the cotangent bundle $T^\ast Q$. We consider the Dirac structure $D_{ \Delta _P}$ induced, via \eqref{def_Dirac},  by the distribution $ \Delta _P$ and the presymplectic form $ \Omega _P:= \pi _{(P, T^*Q)} ^\ast\Omega _{T^*Q}$ on $P$, where $\Omega _{T^*Q}$ is the canonical symplectic form on $ T^*Q$ and $ \pi _{(P, T^*Q)}: P \rightarrow T^*Q$ is the projection defined by $\pi _{(P, T^*Q)}(q,v,p)= (q,p)$. Then, the equations of motion for the nonholonomic system with  $L$ and $\Delta_Q$ can be written in the form of a \textit{Dirac dynamical system} \eqref{DDS} on the Pontryagin bundle as
\begin{equation}\label{DDS_NH} 
\big((\dot q(t), \dot v(t), \dot p(t)) ,\mathbf{d} E(q(t), v(t), p(t)) \big) \in D_{\Delta _P}(q(t), v(t), p(t)),
\end{equation} 
where $E: P \rightarrow \mathbb{R}  $ is the \textit{generalized energy} associated to $L$, defined by $E(q,v,p)= \left\langle p, v\right\rangle - L(q,v)$. The system \eqref{DDS_NH} is equivalent to the implicit differential-algebraic equation
\begin{equation}\label{implicit_NH} 
\left\{
\begin{array}{l}
\displaystyle \vspace{0.2cm} \dot q(t) \in \Delta _Q(q(t)), \qquad v(t)=\dot q(t),\\
\displaystyle \vspace{0.2cm}p(t) = \frac{\partial L}{\partial v} (q(t),v(t)), \qquad  \dot p(t)-  \frac{\partial L}{\partial q}(q(t),v(t)) \in  \Delta _Q(q(t))^\circ,
\end{array} \right.
\end{equation} 
where $ \Delta _Q(q) ^ \circ$ denotes the annihilator of $\Delta _Q(q)$ in $T^*_qQ$.
This system implies the {\it Lagrange-d'Alembert equations} for nonholonomic mechanical systems as
\[
\frac{d}{dt} \frac{\partial L}{\partial v} (q(t),\dot q(t))- \frac{\partial L}{\partial q}(q(t),\dot q(t)) \in  \Delta _Q(q(t))^\circ, \qquad \dot q(t) \in \Delta _Q(q(t)).
\]

\paragraph{The Lagrange-d'Alembert-Pontryagin principle.} For the Dirac dynamical system \eqref{DDS_NH}, there exists an associated variational formulation, in the sense that the critical curves for this principle are exactly the solutions of the system \eqref{implicit_NH}.

This variational formulation is called the \textit{Lagrange-d'Alembert-Pontryagin} principle, which is given by the critical condition for curves $(q(t),v(t),p(t))\in P$ as
\begin{equation}\label{LdAP_linear}
\delta\int_{t_1}^{t_2} \Big[ \big\langle p(t), \dot q(t)\big\rangle  -E(q(t),v(t),p(t))\Big]dt=0,
\end{equation}
where the critical curve $(q(t),v(t),p(t))\in P$ has to satisfy $\dot q(t)\in \Delta_Q(q(t))$ and where the variations considered in \eqref{LdAP_linear} are subject to the conditions $\delta q(t)\in \Delta_Q(q(t))$ and $\delta q(t_1)=\delta q(t_2)=0$.

We can develop the intrinsic expression of the above variational formulation. To do this, defining the 1-form $\Theta_P:=\pi_{(P, T^*Q)}^*\Theta_{T^*Q}$, where $\Theta_{T^*Q}$ is the canonical one-form on $T^*Q$ and writing $\mathrm{x}(t)=(q(t), v(t), p(t))\in P$, it follows that the Lagrange-d'Alembert-Pontryagin principle can be written in the intrinsic form as
\begin{equation}\label{LdAP_theta}
\delta\int_{t_1}^{t_2} \Big[ \left\langle \Theta_P(\mathrm{x}(t)), \dot{\mathrm{x}}(t)\right\rangle  -E(\mathrm{x}(t))\Big]dt=0,
\end{equation}
where the critical curve $\mathrm{x}(t) \in P$ has to satisfy $\dot{\mathrm{x}}(t)\in \Delta_P(\mathrm{x}(t))$ and where the variations considered in \eqref{LdAP_theta} are subject to the conditions $\delta \mathrm{x}(t)\in \Delta_P(\mathrm{x}(t))$ and $T\pi_{P,Q}(\delta \mathrm{x}(t_1))=T\pi_{P,Q}(\delta \mathrm{x}(t_2))=0$. By direct computations, we obtain the following equations:
\begin{equation}\label{intrinsic_LDAPEqn}
\mathbf{i}_{\dot{\mathrm{x}}(t)}\Omega_{P}-\mathbf{d}E(\mathrm{x}(t)) \in 
\Delta_{P}(\mathrm{x}(t))^{\circ},\qquad \dot{\mathrm{x}}(t)\in \Delta_P(\mathrm{x}(t)).
\end{equation}
These equations are clearly equivalent to the Dirac system $\big(\dot{\mathrm{x}}(t), \mathbf{d}E(\mathrm{x}(t))\big) \in D_{\Delta_{P}}(\mathrm{x}(t))$ in \eqref{DDS_NH}, and also are the intrinsic expressions of \eqref{implicit_NH}.
\medskip

As to the details on the variational structures as well as the Dirac dynamical systems, see \cite{YoMa2006a, YoMa2006b}.

\paragraph{Dirac structures in thermodynamics: from adiabatically closed to open systems.} Both the Dirac system formulation and the Lagrange-d'Alembert-Pontryagin variational formulation were developed in \cite{GBYo2018a} for \textit{isolated} thermodynamic systems, based on the variational formulation for nonequilibrium thermodynamics of \cite{GBYo2017a,GBYo2017b}. The key step was the interpretation of the entropy production as a constraint, called a \textit{phenomenological constraint}, as well as the introduction of the concept of \textit{thermodynamic displacements}. This concept allows us to systematically relate the phenomenological constraint to the variational constraint, i.e., the constraints on the variations to be used in the variational condition. A main difficulty in the formulation of Dirac systems for thermodynamics lies in the fact that the phenomenological constraint is \textit{nonlinear}, whereas a Dirac structure provides a fiber-wise linear structure induced from linear constraints (distributions) as in \eqref{def_Dirac}. Nevertheless, the \textit{special relation between the phenomenological and variational constraints} allows for the development of a Dirac system formulation in the case of simple isolated systems, as shown in \cite{GBYo2018a}. This special class of nonlinear constraints is called \textit{nonlinear constraints of thermodynamic type}.

For \textit{open} systems, the exchange of matter between the system and the exterior introduces two additional difficulties at the level of the constraints. First, the constraint becomes now explicitly time-dependent and second, the link between the phenomenological and variational constraints that held in the adiabatically closed case is now broken by additional terms that only depend on the exterior of the system.
Remarkably, both difficulties are simultaneously solved by using the geometric setting of field theories as it applies to the case of \textit{time-dependent mechanics}. In particular, the covariant Pontryagin bundle and the covariant generalized energy must be used instead of the Pontryagin bundle and generalized energy.
In this setting, the time-dependent \textit{nonlinear} constraint associated to the entropy production in the open thermodynamic system can be obtained from the Dirac system associated to a \textit{linear} distribution on the covariant Pontryagin bundle, to which is naturally associated a Dirac structure.

\section{Variational and geometric settings for open systems}\label{Sect_FundSet}

In this section, we review the fundamental setting for open finite dimensional  thermodynamic systems. First we recall in \S\ref{Sect_1st2ndLaws} the first and second laws of thermodynamics and their application to open systems. Then in \S\ref{Section_Vari} we review from \cite{GBYo2018c} the variational formulation for open finite dimensional  thermodynamic systems. This formulation is an extension of the Hamilton principle of classical mechanics. Finally in \S\ref{subsec_gen_setting} we present an abstract setting that  encodes the main features of the variational formulation for open systems since this will be the starting point for the development of the formulation in terms of Dirac structures.

\subsection{The first and second laws for open systems}\label{Sect_1st2ndLaws}

Before going into details on  the fundamental laws in thermodynamics, we need to recall some basic terminology. We denote by $ \boldsymbol{\Sigma} $ a thermodynamic system and by $ \boldsymbol{\Sigma} ^{\rm ext}$ its exterior. The state of the system is completely described by  a set of \textit{state variables}, which may comprise mechanical as well as thermodynamic variables. By definition, \textit{state functions} are given as functions of these state variables. The evolution equations are to be differential equations that determine all the state variables, and hence all the state functions, at any time $t$. We follow the formulation of the two laws as given by \cite{StSc1974}.

\paragraph{The first law of thermodynamics.} 
For every system $ \boldsymbol{\Sigma} $, there exists an \textit{extensive scalar state function} $E$, called {\bfi energy}, which satisfies
\begin{equation}\label{law1_explicit}
\frac{d}{dt} E  =P^{\rm ext}_W+P^{\rm ext}_H+P^{\rm ext}_M,
\end{equation}
where $ P^{\rm ext}_W$ is the power associated to the work done on the system, $P^{\rm ext}_H$ is the power associated to the transfer of heat into the system, and $P^{\rm ext}_M$ is the power associated to the transfer of matter into the system.

Given a thermodynamic system $ \boldsymbol{\Sigma} $, the following terminology is generally adopted:
\begin{itemize}
\item $ \boldsymbol{\Sigma} $ is said to be \textit{closed} if there is no exchange of matter, i.e.,  $P^{\rm ext}_M=0$. When $P^{\rm ext}_M \ne 0$ the system is said to be \textit{open}.
\item 
$ \boldsymbol{\Sigma} $ is said to be \textit{adiabatically closed} if it is closed and there is no heat exchanges, i.e., $P^{\rm ext}_M=P^{\rm ext}_H=0$. 
\item 
$ \boldsymbol{\Sigma} $ is said to be \textit{isolated} if it is adiabatically closed and there is no mechanical power exchange, i.e., $P^{\rm ext}_M=P^{\rm ext}_H=P^{\rm ext}_W=0$.
\end{itemize}

To describe $P_M^{\rm ext}$, let us consider an open system with several ports, $a=1,...,A$, through which matter can flow into or out of the system. We suppose, for simplicity, that the system involves only one chemical species and denote by $N$ the number of moles of this species. The mole balance equation is
\[
\frac{d}{dt}N=\sum_{a=1}^A \mathcal{J}^a,
\]
where $\mathcal{J}^a$ is the molar flow rate \textit{into} the system through the $a$-th port, so that $\mathcal{J}^a>0$ for flow into the system and $\mathcal{J}^a<0$ for flow out of the system.

As matter enters or leaves the system, it carries its internal, potential, as well as kinetic energy. This energy flow rate at the $a$-th port is the product $\mathsf{E}^a\mathcal{J}^a$ of the energy per mole (or molar energy) $\mathsf{E}^a$ and the molar flow rate $\mathcal{J}^a$ at the $a$-th port. In addition, as matter enters or leaves the system it also exerts work on the system that is associated with pushing the species into or out of the system. The associated energy flow rate is given at the $a$-th port by $p^a\mathsf{V}^a\mathcal{J}^a$, where $p^a$ and $\mathsf{V}^a$ denote the pressure and the molar volume of the substance flowing through the $a$-th port respectively. In this case, the power exchange due to the mass transfer is given by
\[
P^{\rm ext}_M= \sum_{a=1}^A\mathcal{J}^a(\mathsf{E}^a+ p^a\mathsf{V}^a).
\]

\noindent {\bf The second law of thermodynamics.}  For every system $ \boldsymbol{\Sigma} $, there exists an \textit{extensive scalar state function} $S$, called {\bfi entropy}, which obeys the following two conditions:
\begin{itemize}
\item[(a)]  Evolution part:\\
If $ \boldsymbol{\Sigma} $ is adiabatically closed, the entropy $S$ is a non-decreasing function with respect to time $t$, i.e., 
\[
\frac{d}{dt} S(t)=I(t)\geq 0,
\]
where $I(t)$ is the {\it entropy production rate} of the system accounting for the irreversibility of internal processes.
\item[(b)] Equilibrium part:\\
If $ \boldsymbol{\Sigma} $ is isolated, as time tends to infinity the entropy tends towards a finite local maximum of the function $S$ over all the thermodynamic states $ \rho $ compatible with the system, i.e., 
\[
\lim_{t \rightarrow +\infty}S(t)= \max_{ \rho \; \text{compatible}}S[\rho ].
\]
\end{itemize}

\paragraph{Finite dimensional simple systems.} A \textit{finite dimensional system} $ \boldsymbol{\Sigma} $, which is also sometimes called a {\it discrete} system, is defined by a collection $ \boldsymbol{\Sigma} =\cup_{A=1}^N \boldsymbol{\Sigma} _A $ of a finite number of interacting simple systems $ \boldsymbol{\Sigma} _A $. Following \cite{StSc1974}, a \textit{simple} system $ \boldsymbol{\Sigma}$ is a macroscopic system for which one (scalar) thermal variable and a finite set of non-thermal variables are sufficient to describe entirely the state of the system. From the second law of thermodynamics, we can always choose the thermal variable as the entropy $S$.

\paragraph{Expression of the entropy production for open systems.} We shall recall the expression of the entropy production for the particular case, in which the system is simple, namely, the system has a single chemical component $N$ in a single compartment with constant volume $V$, and with a single entropy $S$ attributed to the whole system. Also, we assume that there is no heat and work exchanges, except the ones associated to the transfer of matter and ignore all the mechanical effects. We refer to \cite{GBYo2018c} for the expression of the entropy production in the general case.  
Regarding engineering applications and general treatments of open systems, see, for instance, \cite{Fu2010,Sa2006}.

\begin{figure}[h]
\begin{center}
\includegraphics[scale=.70]{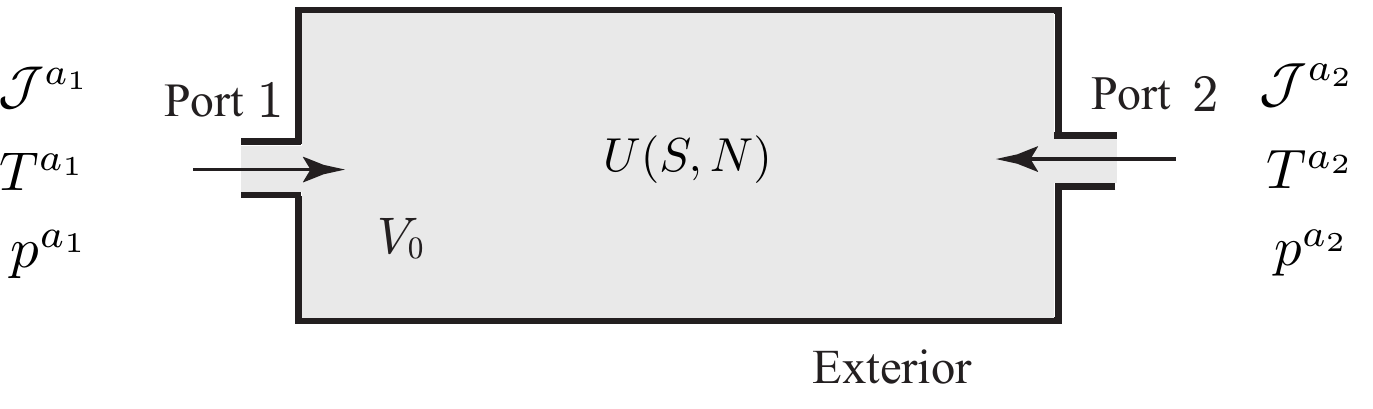}
\caption{A simple open system with two ports}
\label{SimpleOpenSystem}
\end{center}
\end{figure}

In this particular situation, the energy of the system is given by an internal energy $U=U(S,N)$, where $V=V_0$ is constant. The balance of mole and the balance of energy, i.e., the first law (see \eqref{law1_explicit}), are respectively given by
\[
\frac{d}{dt}N=\sum_{a=1}^A \mathcal{J}^a, \quad \frac{d}{dt} U  = \sum_{a=1}^A\mathcal{J}^a(\mathsf{U}^a+ p^a\mathsf{V}^a)=\sum_{a=1}^A\mathcal{J}^a\mathsf{H}^a,
\]
where $\mathsf{H}^a=\mathsf{U}^a+ p^a\mathsf{V}^a$ is the molar enthalpy at the $a$-th port and where $\mathsf{U}^a$, $ p^a$, and $\mathsf{V}^a$ are respectively the molar internal energy, the pressure and the molar volume at the $a$-th port. From these equations and the second law, one obtains the equations for the rate of change of the entropy of the system as

\begin{equation}\label{S_dot_simple}
\frac{d}{dt}S= I+\sum_{a=1}^A \mathsf{S}^a\mathcal{J}^a,
\end{equation}
where $\mathsf{S}^a$ is the molar entropy at the $a$-th port and $I$ is the rate of internal entropy production of the system given by

\begin{equation}\label{I_simple_example}
I= \frac{1}{T}\sum_{a=1}^A \mathcal{J}^a\left(\mathsf{H}^a-T\mathsf{S}^a- \mu \right),
\end{equation}
with $T= \frac{\partial U}{\partial S}$ the temperature and $\mu=\frac{\partial U}{\partial N}$ the chemical potential.
For our variational treatment, it is useful to rewrite the rate of internal entropy production as
\begin{equation}\label{I_simple_example_useful_form}
I= \frac{1}{T}\sum_{a=1}^A \big[\mathcal{J}^a_S(T^a-T)+\mathcal{J}^a(\mu^a- \mu )\big],
\end{equation}
where we used the entropy flow rate $\mathcal{J}^a_S:=\mathsf{S}^a\mathcal{J}^a$ as well as the relation $\mathsf{H}^a=\mathsf{U}^a+ p^a\mathsf{V}^a= \mu^a +T^a\mathsf{S}^a$. The thermodynamic quantities known at the ports are usually the pressure and the temperature $p^a$, $T^a$, from which the other thermodynamic quantities, such as $\mu^a=\mu^a(p^a,T^a)$ or $\mathsf{S}^a=\mathsf{S}^a(p^a,T^a)$ are deduced from the state equations of the gas.
\medskip

The expression of the internal entropy production as rewritten in \eqref{I_simple_example_useful_form} is fundamental for the development of the variational formulation, as we will see below.

\subsection{Variational formulation for open simple systems}\label{Section_Vari}

Now we present the variational formulation for a simple discrete (finite dimensional) open thermodynamic system by following \cite{GBYo2018c}. We shall focus on a simplified situation, namely, the case of an open system with only one entropy variable and one compartment with a single species. We refer to \cite{GBYo2018c} for the more general cases.

\medskip

\paragraph{State variables, Lagrangian, and thermodynamic displacements.} The state variables needed to describe the system are
\[
(q,v,S,N)\in TQ\times \mathbb{R}\times \mathbb{R},
\]
where $Q$ denotes an $n$-dimensional configuration manifold of the mechanical part of the system, $q \in Q$ is the configuration variable and $v\in T_qQ$ is the velocity associated to the mechanical part of the system.  The variable $S$ is the entropy of the system and $N$ is the number of moles of the chemical species.
The Lagrangian is a function defined on the state space, namely,
\[
\mathsf{L}: TQ\times\mathbb{R}  \times \mathbb{R}   \rightarrow \mathbb{R} , \quad (q, v,S, N) \mapsto \mathsf{L}(q, v,S, N),
\]
and is usually given by the kinetic energy minus the internal energy of the system.
We assume that the system has $A$ ports, through which species can flow out or into the system and $B$ heat sources. As above, $\mu^a$ and $T^a$ denote the chemical potential and temperature at the $a$-th port and $T^b$ denotes the temperature of the $b$-th heat source.

\begin{figure}[h]
\begin{center}
\includegraphics[scale=.90]{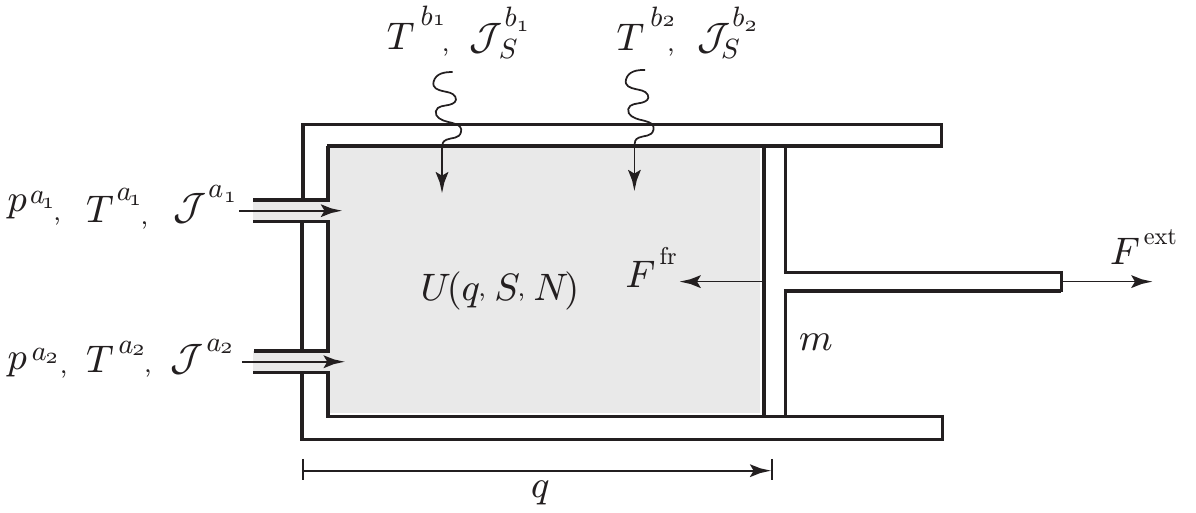}
\caption{A piston with two exterior ports and heat sources}
\label{FluidPiston2}
\end{center}
\end{figure}

An essential ingredient for our variational formulation of thermodynamics is the concept of \textit{thermodynamic displacements} (\cite{GBYo2017a,GBYo2017b,GBYo2018c}). By definition, the \textit{thermodynamic displacement associated with an irreversible process} is given by the primitive in time of a thermodynamic force (or affinity) of the process. For the process of heat transfer, such an affinity is given as the temperature $T$, and hence the thermodynamic displacement is a variable $\Gamma$ such that $\dot{\Gamma}=T$. The variable $\Gamma$ is known as a \textit{thermal displacement}. For the process of mass transfer, the affinity is the chemical potential $\mu$, and hence the thermodynamic displacement is a variable $W$ such that $\dot{W}=\mu$. In addition to these thermodynamic displacements, our variational formulation also involves the variable $\Sigma$, with entropy units, which is defined by the primitive in time of the rate of \textit{internal entropy production} of the system, and therefore $\Sigma$ is distinct from $S$.

\paragraph{Variational formulation for open systems.} With the setting mentioned above, the variational formulation is given as follows.
Find the curves $q(t), S(t), \Gamma(t), \Sigma(t), W(t), N(t)$ which are critical for the \textit{variational condition}
\begin{equation}\label{VCond_open}
\delta \int_{t _1 }^{ t _2} \!\Big[ \mathsf{L}(q, \dot q, S,N) + \dot W N+  \dot{\Gamma }( S- \Sigma )\Big] dt + \int_{t_1}^{t_2}  \left< F^{\rm ext }, \delta q \right> dt =0,
\end{equation}
subject to the \textit{kinematic constraint}
\begin{equation}\label{PC_open}
\frac{\partial \mathsf{L}}{\partial S}\dot \Sigma  =  \left< F^{\rm fr }, \dot q \right>   +\sum_{a=1}^A \Big[\mathcal{J}^a(\dot W- \mu^a) +\mathcal{J}_S^a(\dot \Gamma- T^a)\Big] + \sum_{b=1}^B\mathcal{J}_S^b(\dot \Gamma- T^b),
\end{equation}
and for variations subject to the \textit{variational constraint}
\begin{equation}\label{VC_open}
\frac{\partial \mathsf{L}}{\partial S}\delta \Sigma  =  \left< F^{\rm fr }, \delta q \right>  + \sum_{a=1}^A \Big[\mathcal{J}^a \delta W +\mathcal{J}_S^a \delta \Gamma\Big] + \sum_{b=1}^B\mathcal{J}_S^b \delta \Gamma,
\end{equation}
with $\delta q(t_1)=\delta q(t_2)=0$, $ \delta W(t_1)=\delta W(t_2)=0$, and $ \delta \Gamma(t_1)=\delta \Gamma(t_2)=0$.
\medskip

In the above, the variational constraint \eqref{VC_open} follows from the phenomenological constraint \eqref{PC_open} by formally replacing the time derivatives $\dot\Sigma$, $\dot q$, $\dot W$, $\dot \Gamma$ by the corresponding virtual displacements $\delta\Sigma$, $\delta q$, $\delta W$, $\delta \Gamma$, and by removing all the terms that depend uniquely on the exterior, i.e., the terms $\mathcal{J} ^a\mu ^a$, $\mathcal{J} ^a_S T ^a $, and $ \mathcal{J} ^b_ST ^b $. Such a systematic correspondence between the phenomenological and variational constraints is at the core of the Dirac formulation that we will present below.
We will consider in \S\ref{subsec_gen_setting} a general setting that describes this systematic correspondence between the two constraints.

\medskip

Taking variations of the integral in \eqref{VCond_open}, integrating by parts, and using $ \delta q(t_1)=\delta (t_2)=0$, $\delta W(t_1)=\delta W(t_2)=0$, and $\delta \Gamma(t_1)=\delta \Gamma(t_2)=0$ and using the variational constraint \eqref{VC_open}, we get the following equations:
\begin{equation}\label{conditions_open}
\begin{aligned}
\delta q:&\quad\frac{d}{dt}\frac{\partial \mathsf{L}}{\partial \dot q^{i}}- \frac{\partial \mathsf{L}}{\partial  q^{i}} = F^{\rm fr}_i+F^{\rm ext}_i,\quad i=1,...,n, \\
\delta S:&\quad\dot \Gamma = -\frac{\partial \mathsf{L}}{\partial S}, \\
\delta W:&\quad\dot N= \sum_{a=1}^A\mathcal{J} ^a, \\
\delta N:&\quad\dot W=  -\frac{\partial \mathsf{L}}{\partial N}, \\
\delta \Gamma :&\quad\dot S=\dot \Sigma +\sum_{a=1}^A\mathcal{J}^a_S+\sum_{b=1}^B\mathcal{J}^b_S.
\end{aligned}
\end{equation}
By the second and fourth equations in \eqref{conditions_open}, the variables $\Gamma$ and $W$ are thermodynamic displacements as before. Using \eqref{PC_open}, we get the following system of evolution equations for the curves $q(t)$, $S(t)$, $N(t)$:
\begin{equation}\label{open_system} 
\left\{
\begin{array}{l}
\displaystyle\vspace{0.2cm}\frac{d}{dt}\frac{\partial \mathsf{L}}{\partial \dot q^i}- \frac{\partial \mathsf{L}}{\partial q^i}= F^{\rm fr}_i +F^{\rm ext}_i,\;\;i=1,...,n,\quad\quad\quad  \frac{d}{dt} N= \sum_{a=1}^A \mathcal{J}^a,\\
\displaystyle\vspace{0.2cm}\frac{\partial \mathsf{L}}{\partial S}\Big(\dot S -\sum_{a=1}^A \mathcal{J}_S^a - \sum_{b=1}^B \mathcal{J}_S^b\Big)\\
\displaystyle\quad=  \left< F^{\rm fr }, \dot q \right>
-\sum_{a=1}^A\left[\mathcal{J}^a\Big(\frac{\partial \mathsf{L}}{\partial N}+\mu^a\Big)+\mathcal{J}^a_S\Big(\frac{\partial \mathsf{L}}{\partial S}+ T^a\Big)\right]-\sum_{b=1}^B \mathcal{J}^b_S\Big(\frac{\partial \mathsf{L}}{\partial S}+ T^b\Big).
\end{array} \right.
\end{equation} 
The energy balance for this system is computed as
\[
\frac{d}{dt}E(q, \dot q, S,N)=\underbrace{\left\langle F^{\rm ext}, \dot q\right\rangle \textcolor{white}{\sum_{b=1}^B\hspace{-0.5cm}}}_{=P^{\rm ext}_W} \;+\; \underbrace{\sum_{b=1}^B \mathcal{J}^b_S T^b}_{=P^{\rm ext}_H} \; + \; \underbrace{\sum_{a=1}^A (\mathcal{J}^a\mu^a + \mathcal{J}_S^a T^a)}_{=P^{\rm ext}_M},
\]
where the energy $E(q, \dot q, S,N)$ is given by $E=\left<\frac{\partial \mathsf{L}}{\partial \dot{q}}, \dot{q}\right>-\mathsf{L}$.
From the last equation in \eqref{open_system}, the rate of entropy equation of the system is found as
\begin{equation}\label{entropy_open}
\dot S=I + \sum_{a=1}^A \mathcal{J}^a_S + \sum_{b=1}^B \mathcal{J}^b_S,
\end{equation}
where $I$ is the rate of internal entropy production given by
\[
I= \underbrace{-\frac{1}{T}\left< F^{\rm fr }, \dot q \right>\textcolor{white}{\sum_{b=1}^B\hspace{-0.5cm}}}_{\text{mechanical friction}} +\underbrace{\frac{1}{T} \sum_{a=1}^A\left[\mathcal{J}^a\Big(\mu ^ a- \mu\Big)+\mathcal{J}^a_S\Big( T^a - T\Big)\right]}_{\text{mixing of matter flowing into the system}} \; + \; \underbrace{\frac{1}{T}\sum_{b=1}^B \mathcal{J}^b_S\Big(T^b- T\Big)}_{\text{heating}}.
\]
From the last equation \eqref{conditions_open} and \eqref{entropy_open} it follows that $\dot\Sigma =  I$ is the rate of internal entropy production.
The second and third terms in \eqref{entropy_open} represent the entropy flow rate into the system associated to the ports and the heat sources. The second law requires $I\geq 0$, whereas the sign of the rate of entropy flow into the system is arbitrary.

\subsection{A general setting for open thermodynamic systems}\label{subsec_gen_setting}

In order to formulate open thermodynamic systems in terms of Dirac structures, it is useful to develop a general variational setting that encodes the main features of the variational formulation \eqref{VCond_open}--\eqref{VC_open}. Details on how this setting applies to the thermodynamics of open systems are clarified in the last paragraph.

\paragraph{General setting.} Let us denote by $\mathcal{Q}$ an $n$-dimensional configuration manifold of the system. An element $x \in \mathcal{Q}$ represents all of the variables involved in the system, namely, not only mechanical variables but also the thermodynamic variables. The phenomenological constraint for open systems can be written in general form as
\begin{equation}\label{General_constraint}
\sum_{i=1}^n A^r_i(t,x,\dot x)\dot x^i+B^r(t,x,\dot x)=0,\quad \text{for $r=1,...,m<n$},
\end{equation}
where $A^r:\mathbb{R}\times T \mathcal{Q}\rightarrow T^*\mathcal{Q}$ is a fiber preserving map, i.e., $A^r(t,x,v)\in T^*_x\mathcal{Q}$, for all $(t,x,v)\in\mathbb{R}\times T\mathcal{Q}$, and $B^r: \mathbb{R}\times T \mathcal{Q}\rightarrow\mathbb{R}$, for all $r=1,...,m$. Note that the phenomenological constraint \eqref{General_constraint} is \textit{nonlinear} and \textit{time-dependent}.

Consider a Lagrangian $L:\mathbb{R}\times T\mathcal{Q}\rightarrow\mathbb{R}$ and an external force $\mathcal{F}^{\rm ext}: \mathbb{R}\times T \mathcal{Q}\rightarrow T^*\mathcal{Q}$, with $\mathcal{F}^{\rm ext}(t,x,v)\in T^*_x\mathcal{Q}$, for all $(t,x,v)\in\mathbb{R}\times T\mathcal{Q}$. Note that in this general setting we allow both $L$ and $\mathcal{F}^{\rm ext}$ to be time-dependent, although this is not the case in \eqref{VCond_open}.

\paragraph{Variational formulation.} Given $A^r$, $B^r$, $L$, and $\mathcal{F}^{\rm ext}$ as above, consider the generalized version of the \textit{Lagrange-d'Alembert principle} as follows: Find the curve $x(t)\in\mathcal{Q}$ critical for the \textit{variational condition}
\begin{equation}\label{Vcond_abstract}
\delta \int_{t _1 }^{t _2 }  L(t,x,\dot{x}) {\rm d} t+\int_{t _1 }^{t _2 } \left\langle \mathcal{F}^{\rm ext}(t, x, \dot x), \delta x \right\rangle {\rm d}t=0,
\end{equation}
subject to the \textit{kinematic constraint}
\begin{equation}\label{PC_abstract}
\sum_{i=1}^nA_i^r(t,x,\dot x )\dot x ^ i  +B^ r (t,x,\dot x )=0,\;\; r=1,...,m<n,
\end{equation}
and for variations $\delta x$ subject to the \textit{variational constraint}
\begin{equation}\label{VC_abstract}
\sum_{i=1}^n A_i^r(t,x,\dot x )\delta x^i =0,\;\; r=1,...,m,
\end{equation}
with $\delta x^i (t_1)=\delta x^i (t_ 2)=0$.

\paragraph{The generalized Lagrange-d'Alembert equations.} By direct computations, using Lagrange multipliers $\lambda_r$, $r=1,...,m<n$, a curve $x(t)\in\mathcal{Q}$ is critical for the variational formulation \eqref{Vcond_abstract}--\eqref{VC_abstract} if and only if it is a solution of the following Lagrange-d'Alembert equations with time-dependent and nonlinear nonholonomic constraints and with external forces:
\begin{equation}\label{evo_eqn_nh1}
\left\{
\begin{array}{l}
\vspace{0.2cm}\displaystyle\frac{d}{dt}\frac{\partial L}{\partial \dot{x}^{i}}(t,x,\dot{x})- \frac{\partial L}{\partial x^{i}}(t,x,\dot{x})=\sum_{r=1}^m \lambda_{r}A_{i}^{r}(t,x,\dot{x})+\mathcal{F}^{\rm ext}_{i}(t, x, \dot x),\\
\displaystyle \sum_{i=1}^n A_i^r(t,x,\dot x )\dot x ^ i  +B^ r (t,x,\dot x )=0.
\end{array}\right.
\end{equation}

Associated to the Lagrangian $L(t, x,v)$, we define the energy $E_L:\mathbb{R}\times T\mathcal{Q}\rightarrow \mathbb{R}$ by
\begin{equation}\label{E_L}
E_L(t,x,v)=\left\langle\frac{\partial L}{\partial v}(t,x,v ), v\right\rangle-L(t,x,v).
\end{equation}
On the solutions of the equations \eqref{evo_eqn_nh1}, we have the {\it energy balance equation}
\begin{equation}\label{energy_t}
\frac{d}{dt}E_L(t,x,\dot x)=\left\langle \mathcal{F}^{\rm ext} (t, x, \dot x),\dot x\right\rangle-\sum_{r=1}^m \lambda_{r}B^r(t,x,\dot x )-\frac{\partial L}{\partial t}(t,x,\dot x ).
\end{equation}

\paragraph{Application of the general setting to open systems.} The variational setting for open systems given in \S\ref{Section_Vari} can be obtained from the general setting above by choosing the configuration manifold
\[
\mathcal{Q}= Q\times \mathbb{R}^5\ni x=(q,S,N,\Gamma, W, \Sigma),
\]
where $Q$ is the configuration manifold of the mechanical part of the system.
The Lagrangian $L:\mathbb{R}\times T\mathcal{Q}\rightarrow \mathbb{R}$ is time-independent and given by
\begin{equation}\label{choice_L}
L(t,x,\dot x)=\mathsf{L}(q, \dot q, S,N) + \dot W N+  \dot{\Gamma }( S- \Sigma ).
\end{equation}
To describe the constraints, we choose the coefficients $A^r_i(t,x,\dot x)$ and $B^r_i(t,x,\dot x)$ such that
\begin{align*}
\sum_{i=1}^nA^r_i(t,x,\dot x)\delta x^i&= - \frac{\partial \mathsf{L}}{\partial S} \delta \Sigma  +  \left< F^{\rm fr }, \delta q \right>   +\sum_{a=1}^A \Big[\mathcal{J}^a\delta W +\mathcal{J}_S^a\delta \Gamma\Big] + \sum_{b=1}^B\mathcal{J}_S^b\delta \Gamma,\\
B^r_i(t,x,\dot x)&=- \sum_{a=1}^A \Big[\mathcal{J}^a \mu^a- \mathcal{J}_S^aT^a(t)\Big] - \sum_{b=1}^B\mathcal{J}_S^bT^b.
\end{align*}
Note that here $m=1$, so that the exponent $r=1$ can be ignored.
With these choices, the abstract kinematic constraint \eqref{PC_abstract} and variational constraint \eqref{VC_abstract} recover the kinematic constraint \eqref{PC_open} and variational constraint \eqref{VC_open} required in open thermodynamic systems. 
\medskip

It is important to observe that the molar flow rates $\mathcal{J}^a$,  the entropy flow rates $\mathcal{J}^a_S$, $\mathcal{J}^b_S$, and the temperatures and chemical potentials $T^a$, $T^b$, $\mu^a$ at the ports may in general be explicit functions of time and some variables of the system, i.e., $\mathcal{J}^a=\mathcal{J}^a(t,x,\dot x)$, $T^a=T^a(t,x,\dot x)$, for instance.
This is why it is important to allow the constraints to be explicitly dependent on time in our general setting.
Although in general the Lagrangian, as shown in \eqref{choice_L}, is not explicitly dependent on time in our examples, we dare to regard the Lagrangian as a time-dependent one, since the time-dependence of the constraint demands to consider the geometric setting for time-dependent systems.

\section{Dirac structures for time-dependent nonholonomic systems of thermodynamic type}\label{Section_3}

In this section let us first consider two geometric objects, $C_V$ and $C_K$, namely, the variational and kinematic constraints used in  \eqref{PC_abstract} and \eqref{VC_abstract}. We clarify the link between $C_V$ and $C_K$ that specifically appear in open thermodynamic systems, which are called the constraints of \textit{thermodynamic type}. In this situation, we construct from $C_V$ a distribution on the covariant Pontryagin bundle. The Dirac structure is induced from this distribution and from a presymplectic form on the covariant Pontryagin bundle.  Then, we can formulate the Dirac dynamical system associated to this Dirac structure and to the covariant generalized energy, and obtain the equations of evolution governed by this system. Finally, we show that there exists an associated variational formulation whose critical condition is exactly the condition given by the Dirac dynamical system. Furthermore, two other Dirac system formulations are presented, which are respectively associated with the Lagrangian and the Hamiltonian. These Dirac systems, called Lagrange-Dirac and Hamilton-Dirac systems, are based on the Dirac structure induced on the cotangent bundle rather than the covariant Pontryagin bundle.

\subsection{Geometric setting for time-dependent constraints of thermodynamic type}

For the definition of the Dirac structure below, it is useful to define two geometric objects $C_V$ and $C_K$ associated with the variational and kinematic constraints used in \eqref{PC_abstract} and \eqref{VC_abstract}. Since these constraints depend explicitly on time, it is useful to establish a geometric setting that consistently includes time.

Given the configuration manifold $\mathcal{Q}$ considered above, we define the \textit{extended configuration manifold}
\[
Y:=\mathbb{R}\times \mathcal{Q}\ni (t,x),
\]
seen as a trivial vector bundle over $\mathbb{R}$, namely, $Y=\mathbb{R}\times\mathcal{Q}\rightarrow \mathbb{R}$, $(t,x)\mapsto t$. This is known as the geometric setting of  time-dependent mechanics in the context of classical field theories, as in Remark \ref{field_theory} below. In this setting, $Y\rightarrow \mathbb{R}$ is the configuration bundle of the field theory. This setting will guide the developments made in this Section.

\paragraph{Time-dependent variational and kinematic constraints.} 
Consider the vector bundle $(\mathbb{R}\times T\mathcal{Q})\times_Y T Y\rightarrow Y$ over $Y$ whose vector fiber at $y=(t,x) \in Y$ is given by $T_x\mathcal{Q} \times T_{(t,x)}Y=T_x\mathcal{Q} \times (\mathbb{R}\times  T_x\mathcal{Q})$. So an element in the fiber at each $y=(t,x)$ is denoted $(v,\delta t, \delta x)$. By definition a \textit{variational constraint} is a subset
\[
C_V \subset (\mathbb{R}\times T\mathcal{Q})\times_Y T Y,
\]
such that $C_V(t,x,v)$, defined by
\[
C_V(t,x,v):=C_V\cap \left(\{(t,x,v)\}\times T_{(t,x)}Y\right),
\]
is a vector subspace of $T_{(t,x)}Y$, for all $(t,x,v)\in \mathbb{R}\times T\mathcal{Q}$. A \textit{kinematic constraint} is by definition a submanifold
\[
C_K\subset TY.
\]
This setting extends to the time-dependent case, the setting considered in \cite{CeIbdLdD2004} for mechanical systems, in which case $C_V\subset T\mathcal{Q}\times_\mathcal{Q} T\mathcal{Q}$ and $C_K\subset T\mathcal{Q}$.

\paragraph{Constraints of thermodynamic type.} Given $C_V$ and $C_K$ without any specific relation between them, one can always develop a variational formulation based on the generalized Lagrange-d'Alembert principle. However, in general we cannot establish a Dirac structure in such a general situation. For the case of thermodynamic systems, it is remarkable that there is a specific relation between $C_V$ and $C_K$, which directly follows from the variational setting developed as in \cite{GBYo2018c}, and allows for a formulation in terms of Dirac structures. This relation between $C_V$ and $C_K$ is stated as follows: 

\begin{definition}\label{def_thermotype} A variational constraint $C_V\subset (\mathbb{R}\times T\mathcal{Q})\times_Y T Y$ and a kinematic constraint $C_K\subset TY$ are called of {\bfi thermodynamic type} if $C_K$ is defined in terms of $C_V$ as follows:
\begin{equation}\label{CK_CV}
C_{K}=\big\{(t,x,\dot t,\dot x)\in TY \mid (t,x,\dot t,\dot x) \in C_{V}(t,x,\dot x) \big\} \subset TY.
\end{equation}
\end{definition}

For instance, the variational constraint $C_V$ associated to the functions $A^r:\mathbb{R}\times T \mathcal{Q}\rightarrow T^*\mathcal{Q}$ and $B^r: \mathbb{R}\times T \mathcal{Q}\rightarrow\mathbb{R}$, $r=1,...,m$, considered in \S\ref{subsec_gen_setting}, is given as follows
\begin{equation}\label{CV} 
\begin{aligned}
C_{V}&=\big\{(t, x, v,  \delta{t}, \delta{x}) \in (\mathbb{R} \times T\mathcal{Q}) \times_{Y} TY \mid \\
& \hspace{2cm} A_{i}^{r}(t,x,v)\delta x^{i} +B^{r}(t,x,v)\delta{t}=0,\; r=1,...,m\big\}.
\end{aligned}
\end{equation}
Then, the associated kinematic constraint $C_K$ of thermodynamic type defined in \eqref{CK_CV} reads
\begin{equation}\label{CK} 
C_{K}=\big\{(t,x,\dot t ,\dot x) \in TY \mid  A_{i}^{r}(t,x, \dot x) \dot x^{i} +B^{r}(t,x, \dot x)\dot{t}=0,\; r=1,...,m \big\}.
\end{equation}
In the above, we employed Einstein's summation convention, which will be used in the following sections unless otherwise stated.

\subsection{Time-dependent Dirac systems on the covariant Pontryagin bundle}\label{TDCPB}

Recall that in mechanics with a configuration manifold $Q$, the Pontryagin bundle is the vector bundle over $Q$ defined as the Whitney sum of the tangent and cotangent bundles of $Q$, i.e.,
\begin{equation}\label{Pontryagin_mechanics}
P= TQ\oplus T^*Q\rightarrow Q.
\end{equation}
The Pontryagin bundle is the natural object on which the generalized energy is defined. Namely, for a given Lagrangian $L:TQ\rightarrow \mathbb{R}$,
the \textit{generalized energy} $E: P\rightarrow\mathbb{R}$ is defined by
\[
E(q,v,p):= \langle p, v\rangle - L(q,v).
\]

\paragraph{Covariant Pontryagin bundle and generalized energy.}
The analogue to the Pontryagin bundle \eqref{Pontryagin_mechanics} for time-dependent mechanics with an extended configuration manifold $Y=\mathbb{R}\times \mathcal{Q}$ is the \textit{covariant Pontryagin bundle} given by
\begin{equation}\label{Pontryagin_time_mech}
\pi_{(\mathcal{P},Y)}: \mathcal{P}= (\mathbb{R} \times T\mathcal{Q}) \times_{Y} T^\ast Y \rightarrow Y=\mathbb{R}\times\mathcal{Q}.
\end{equation}
See Remark \ref{field_theory} for the justification of this definition. For the covariant Pontryagin bundle, an element  in the fiber  at $y=(t,x) \in Y$ is denoted by $(v,\mathsf{p}, p)$. Given a Lagrangian $L: \mathbb{R}\times T \mathcal{Q}\rightarrow\mathbb{R}$, the \textit{covariant generalized energy} $\mathcal{E}: \mathcal{P}\rightarrow\mathbb{R}$ is defined on $\mathcal{P}$ as
\begin{equation}\label{CovariantEnergy}
\mathcal{E}(t,x,v,\mathsf{p}, p)= \mathsf{p} + \langle p, v\rangle -  L(t,x,v).
\end{equation}
We will also define the \textit{generalized energy} $E: \mathbb{R}\times (T\mathcal{Q}\oplus T^*\mathcal{Q})\rightarrow\mathbb{R}$ by
\begin{equation}\label{gen_energy}
E(t,x,v,p)= \langle p, v\rangle - L(t,x,v).
\end{equation}

\begin{remark}[Geometric setting for field theory and time-dependent mechanics]\label{field_theory}{\rm We shall now comment on the compatibility of the definitions \eqref{Pontryagin_time_mech} and \eqref{CovariantEnergy} with the general definitions of the covariant Pontryagin bundle and covariant generalized energy in field theories, see \cite{GIMM1997}, for the details. To describe field theories, one starts with a fiber bundle, $\pi_{(Y,X)}:Y \rightarrow X$, the \textit{configuration bundle}, whose sections are the fields of the theory. The analogue to the tangent bundle of mechanics is the first jet bundle $J^1Y\rightarrow Y$, whose fiber at $y\in Y$ is the affine space $J^1_yY= \{\gamma_y \in L(T_xX,T_yY)\mid T\pi_{(Y,X)}\circ \gamma_y= {\rm Id}_{T_xX}\}$, where $x=\pi_{(Y,X)}(y)$. The Lagrangian density of the theory is a map
\begin{equation}\label{Lagr_density}
\mathscr{L}:J^1Y\rightarrow \Lambda^{n+1}X,
\end{equation}
covering $\pi_{(Y,X)}:Y \rightarrow X$, where $X$ is an oriented manifold with $\operatorname{dim}X=n+1$ and $\Lambda^{n+1}X$ denotes the bundle of $(n+1)$-form over $X$. The analogue to the cotangent bundle of mechanics is the dual jet bundle $J^1Y^\star\rightarrow Y$, whose fiber at $y\in Y$ is the space of affine maps from $J^1_yY$ to $\Lambda^{n+1}X$, i.e., $J^1_yY^\star= \operatorname{Aff}(J^1_yY, \Lambda^{n+1}_xX)$. The field theoretic analogue to \eqref{Pontryagin_mechanics} is thus the covariant Pontryagin bundle defined as
\begin{equation}\label{Pontryagin_fields}
\mathcal{P}= J^1Y\times _Y J^1Y^\star\rightarrow Y.
\end{equation}
The \textit{covariant generalized energy density} associated to $\mathscr{L}$ is defined on $\mathcal{P}$ as
\[
\mathscr{E}(\gamma,z)= \langle z,\gamma\rangle - \mathscr{L}(\gamma),
\]
where $\gamma\in J^1_yY$ and $z\in J^1_yY^\star$. In local coordinates $\gamma=(x^\mu, y^i, v^i_\mu)$, $z=(x^\mu, y^i, \mathsf{p}, p^\mu_i)$ and the covariant generalized energy density reads
\begin{equation}\label{cov_gen_E_local}
\mathscr{E}(x^\mu, y^i, v^i_\mu, \mathsf{p}, p^\mu_i)= (\mathsf{p} + p^\mu_iv^i_\mu ) d^{n+1}x- \mathscr{L}(x^\mu, y^i, v^i_\mu).
\end{equation}

It is well known that time-dependent mechanics can be geometrically formulated as a field theory. In this case, we have $X=\mathbb{R}$ and the configuration bundle is the trivial bundle $Y=\mathbb{R}\times\mathcal{Q}\rightarrow\mathbb{R}$. We have the canonical identifications $J^1Y\cong \mathbb{R}\times T\mathcal{Q}$ and $J^1Y^\star \cong T^*(\mathbb{R}\times \mathcal{Q})$. Therefore, in this case the general definition of the covariant Pontryagin bundle in \eqref{Pontryagin_fields} does recover \eqref{Pontryagin_time_mech}, since $J^1Y\times _Y J^1Y^\star \cong (\mathbb{R}\times T\mathcal{Q})\times_Y T^*Y$, when $Y=\mathbb{R}\times \mathcal{Q}$.

The Lagrangian density \eqref{Lagr_density} is in this case a map $\mathscr{L}:\mathbb{R}\times T\mathcal{Q} \rightarrow\Lambda^1\mathbb{R}$. The link with the Lagrangian $L:\mathbb{R}\times T\mathcal{Q}\rightarrow\mathbb{R}$ considered above is $\mathscr{L}(t,x,v)=L(t,x,v)dt$. Note that the definition of $L$ in terms of $\mathscr{L}$ depends on the chosen parameterization of time, i.e., of $t\in \mathbb{R}$.

When $Y=\mathbb{R}\times\mathcal{Q}$, by using \eqref{cov_gen_E_local} we see that the covariant generalized energy density $\mathscr{E}$ is related to the covariant generalized energy \eqref{CovariantEnergy} as $\mathscr{E}(t,x,v,\mathsf{p}, p)=\mathcal{E}(t,x,v,\mathsf{p}, p) dt$. Here again, this relation depends on the chosen parameterization of time.
}
\end{remark}

\paragraph{Distribution induced on the covariant Pontryagin bundle.} From a given variational constraint $C_V\subset (\mathbb{R}\times T\mathcal{Q})\times_Y T Y$, we define the induced distribution $\Delta_\mathcal{P}$ on the covariant Pontryagin bundle defined as
\begin{equation}\label{Delta_P}
\Delta_{\mathcal{P}}(t,x,v,\mathsf{p}, p):=\big(T_{(t,x,v,\mathsf{p}, p)}\pi_{(\mathcal{P},Y)}\big)^{-1}(C_{V}(t, x, v)) \subset T_{(t,x,v,\mathsf{p}, p)}\mathcal{P}.
\end{equation}
If $C_V$ is given as in \eqref{CV}, then the distribution $\Delta_\mathcal{P}$ is given by
\begin{equation}\label{delta_P}
\begin{aligned}
\Delta_{\mathcal{P}}(t,x,v,\mathsf{p}, p)&=\big\{(\delta{t},\delta{x},\delta{v},\delta{\mathsf{p}}, \delta{p}) \in T_{(t,x,v,\mathsf{p}, p)}\mathcal{P} \mid  \\ 
& \hspace{2cm} A_{i}^{r}(t,x,v)\delta{x}^{i} +B^{r}(t,x,v)\delta{t}=0,\; r=1,...,m\big\}.
\end{aligned} 
\end{equation}

\paragraph{Dirac structures on the covariant Pontryagin bundle.} Consider the canonical symplectic form on $T^*Y$ given by $\Omega_{T^*Y}=-\mathbf{d}\Theta_{T^*Y}$, where $\Theta_{T^*Y}$ is the canonical one-form on $T^*Y$. In local coordinates, we have $\Theta_{T^*Y}=p_{i}dx^{i}+\mathsf{p}dt$ and $\Omega_{T^*Y}=dx^{i} \wedge dp_{i} + dt \wedge d\mathsf{p}$. Using the projection $\pi_{(\mathcal{P}, T^{\ast}Y)}:\mathcal{P} \rightarrow T^{\ast}Y$, $(t,x,v,\mathsf{p}, p)\mapsto(t,x,\mathsf{p}, p)$ onto $T^{\ast}Y$, we get the presymplectic form on the covariant Pontryagin bundle given by
\begin{equation}\label{presym_P}
\Omega_{\mathcal{P}}=\pi_{(\mathcal{P}, T^{\ast}Y)}^{\ast}\Omega_{T^{\ast}Y}.
\end{equation}
Its local expression is $\Omega_{\mathcal{P}}=dx^{i} \wedge dp_{i} + dt \wedge d{\sf p}$.

Given the distribution $\Delta_{\mathcal{P}}$ in \eqref{Delta_P} and the presymplectic form $\Omega_{\mathcal{P}}$ in \eqref{presym_P}, we consider the Dirac structure $D_{\Delta_{\mathcal{P}}}$ on $\mathcal{P}$ defined as in \eqref{def_Dirac} by, for all $\mathrm{x}\in \mathcal{P}$,
\begin{equation}\label{Dirac_P}
\begin{aligned}
D_{\Delta_{\mathcal{P}}}(\mathrm{x})&=\big\{(\mathfrak{u}_{\mathrm{x}}, \mathfrak{a}_{\mathrm{x}}) \in T_{\mathrm{x}}\mathcal{P} \times T_{\mathrm{x}}^{\ast}\mathcal{P} \mid \mathfrak{u}_{x} \in \Delta_{\mathcal{P}}(x),  \\
& \hspace{3cm} \left<\mathfrak{a}_{\mathrm{x}}, \mathfrak{v}_{\mathrm{x}}\right>= \Omega_{\mathcal{P}}(\mathrm{x})(\mathfrak{u}_{\mathrm{x}}, \mathfrak{v}_{\mathrm{x}}),\;\forall \;\mathfrak{v}_{\mathrm{\mathrm{x}}} \in \Delta_{\mathcal{P}}(\mathrm{x})\big\}.
\end{aligned}
\end{equation}

We now show explicitly this Dirac structure for the distribution given in \eqref{delta_P}. Writing $\mathrm{x}=(t,x,v,\mathsf{p}, p) \in \mathcal{P}$, $\mathfrak{u}_{\mathrm{x}}=(u_t, u_x,u_v, u_{\mathsf{p}}, u_p)\in  T_{\mathrm{x}}\mathcal{P}$, $\mathfrak{v}_{\mathrm{x}}=( \delta{t}, \delta x, \delta{v}, \delta{\mathsf{p}}, \delta{p}) \in T_{\mathrm{x}}\mathcal{P}$, and $\mathfrak{a}_{\mathrm{x}}=(\pi, \alpha, \beta, \gamma, w) \in T^{\ast}_{\mathrm{x}}\mathcal{P}$, the condition $(\mathfrak{u}_{\mathrm{x}}, \mathfrak{a}_{\mathrm{x}}) \in D_{\Delta_{\mathcal{P}}}(\mathrm{x})$ reads
\begin{equation}\label{local_DiracCond_P}
\begin{aligned}
&u_x=w, \quad u_t=\gamma,\quad \beta=0,\\[2mm]
&(t,x,u_t,u_x) \in C_{V}(t,x,v), \quad(u_{\mathsf{p}}+\pi, u_p+\alpha) \in C_{V}(t,x,v)^{\circ},
\end{aligned}
\end{equation}
where $C_{V}(t,x,v)^{\circ} \subset T^{\ast}_{(t,x)}Y$ denotes the annihilator of $C_{V}(t,x,v) \subset T_{(t,x)}Y$, i.e.,
\[
C_{V}(t,x,v)^{\circ}=\big\{ \mathsf{a} \in T^{\ast}_{(t,x)}Y \mid \left<\mathsf{a},\eta \right>=0,\; \forall\;\eta\in C_{V}(t,x,v)  \big\}.
\]
By using \eqref{CV}, we get the annihilator
\[
C_{V}(t,x,v)^{\circ}=\{(t,x, \pi, \alpha)\in T^*_{(t,x)}Y\mid \pi=\lambda_rB^{r}(t,x,v) , \;\alpha=\lambda_r A_{i}^{r}(t,x,v),\;\lambda^r\in\mathbb{R}\}.
\]
Thus, the coordinate expressions in \eqref{local_DiracCond_P} are
\begin{equation}\label{local_Dirac_P}
\begin{aligned}
&u_x^i=w^{i}, \quad u_t=\gamma,\quad \beta_{i}=0,\quad A_{i}^{r}(t,x,v)u_x^{i} +B^{r}(t,x,v)u_t=0,\\[2mm]
& u_{\mathsf{p}}+\pi =\lambda_{r}B^{r}(t,x,v), \quad(u_p)_{i}+\alpha_{i}=\lambda_{r} A^{r}_{i}(t,x,v), \; r=1,...,m.
\end{aligned}
\end{equation}

\paragraph{Dirac dynamical systems on the covariant Pontryagin bundle.} Given a variational constraint $C_V$, consider the induced distribution $\Delta_\mathcal{P}$ on $\mathcal{P}$ defined in \eqref{Delta_P} and the Dirac structure $D_{\Delta_\mathcal{P}}$ defined in \eqref{Dirac_P}. Given also a time-dependent Lagrangian $L:\mathbb{R} \times T\mathcal{Q}\rightarrow \mathbb{R}$, consider the associated covariant generalized energy $\mathcal{E}$ defined in \eqref{CovariantEnergy}.

\begin{definition}
Given $\Delta_\mathcal{P}$ and $L$ as above, the associated {\bfi time-dependent Dirac dynamical system} for a curve of the form
\begin{equation}\label{curve_section}
\mathrm{x}(t)=(t,x(t),v(t),\mathsf{p}(t), p(t)) \in \mathcal{P}
\end{equation}
on the covariant Pontryagin bundle is
\begin{equation}\label{DiracDyn_P}
\big(\dot{\mathrm{x}}(t), \mathbf{d}\mathcal{E}({\mathrm{x}(t)})\big) \in D_{\Delta_{\mathcal{P}}}({\mathrm{x}(t)}).
\end{equation}
\end{definition}

\begin{remark}\label{curve_section_remark}{\rm
Note that the curve $\mathrm{x}(t)\in\mathcal{P}$ in \eqref{curve_section} is not an arbitrary curve in $\mathcal{P}$ since its first component is $t$. In the language of field theory, it is a \textit{section} of the covariant Pontryagin bundle seen as a bundle over $\mathbb{R}$.}
\end{remark}

From the expression \eqref{Dirac_P}, we have the equivalence
\[
\big(\dot{\mathrm{x}}, \mathbf{d}\mathcal{E}({\mathrm{x}})\big) \in D_{\Delta_{\mathcal{P}}}({\mathrm{x}})\quad\Leftrightarrow\quad \mathbf{i}_{\dot{\mathrm{x}}}\Omega_{\mathcal{P}}-\mathbf{d}\mathcal{E}(\mathrm{x}) \in \Delta_{\mathcal{P}}(\mathrm{x})^{\circ},\;\; \dot{\mathrm{x}} \in \Delta_{\mathcal{P}}(\mathrm{x}).
\]
Now, let us explicitly compute the equations of motion of the Dirac dynamical system. The differential of $\mathcal{E}$ is given by
\[
\mathbf{d}\mathcal{E}(t,x,v,\mathsf{p},p)=\left( -\frac{\partial L}{\partial t}, -\frac{\partial L}{\partial x}, p-\frac{\partial L}{\partial v}, 1, v \right).
\]
Therefore, using the expression \eqref{local_DiracCond_P} of the Dirac structure, the Dirac dynamical system \eqref{DiracDyn_P} gives following the conditions on the curve $\mathrm{x}(t)\in\mathcal{P}$, 
\begin{equation}\label{loc_TimeDepDiracSys_P}
\begin{aligned}
&\dot x=v, \qquad \dot t=1,\qquad  p=\frac{\partial L}{\partial v},\\
&(t,x,\dot t,\dot{x}) \in C_{V}(t,x,v),\qquad\left(\dot{\mathsf{p}}-\frac{\partial L}{\partial t}, \dot p-\frac{\partial L}{\partial x}\right) \in C_{V}(t,x,v)^{\circ}.
\end{aligned}
\end{equation}
By using the local expressions in \eqref{local_Dirac_P}, we get the Dirac dynamical system in the following form:
\begin{equation}\label{local_Dirac_DS}
\left\{
\begin{array}{l}
\vspace{0.2cm}\displaystyle\dot x^i = v^i , \qquad  \dot t=1,\qquad p_i-\frac{\partial L}{\partial v^i}=0,\quad i=1,...,n,\\
\vspace{0.2cm}\displaystyle A_{i}^{r}(t,x,v)\dot q^{i}+B^{r}(t,x,v)=0, \quad r=1,...,m,\\
\displaystyle\dot p_i-\frac{\partial L}{\partial x^i}= \lambda_{r} A^{r}(t,x,v), \qquad \dot{\mathsf{p}}-\frac{\partial L}{\partial t} = \lambda_{r}B^{r}(t,x,v).
\end{array}\right.
\end{equation}

Since the second equation in \eqref{local_Dirac_DS} is always satisfied, and since the last equation in \eqref{local_Dirac_DS} can be solved apart from the others (as an output equation), \eqref{local_Dirac_DS} induces the evolution equations
\begin{equation}\label{local_Dirac_Eqn}
\left\{
\begin{array}{l}
\displaystyle\vspace{0.2cm}\dot{x}^i=v^i, \qquad p_i=\frac{\partial L}{\partial v^i}(t,x,v),\qquad \dot{p}_i-\frac{\partial L}{\partial x^i}(t,x,v)=\lambda_{r} A^{r}(t,x,v),\\
\displaystyle A_{i}^{r}(t,x,v)\dot{x}^{i} +B^{r}(t,x,v)=0,
\end{array}
\right.
\end{equation}
for the curve $(x(t),v(t),p(t)) \in T\mathcal{Q} \oplus T^{\ast}\mathcal{Q}$.
Finally, system \eqref{local_Dirac_Eqn} yields the following equations for the curve $x(t)\in \mathcal{Q}$:
\begin{equation}\label{local_Dirac_Eqn_final}
\left\{
\begin{array}{l}
\displaystyle\vspace{0.2cm}\frac{d}{dt}\frac{\partial L}{\partial \dot x^i}-\frac{\partial L}{\partial x^{i}}(t,x,\dot x)=\lambda_{r} A^{r}_{i}(t,x,\dot x),\\
\displaystyle A_{i}^{r}(t,x,\dot x)\dot{x}^{i} +B^{r}(t,x,\dot x)=0,
\end{array}
\right.
\end{equation}
which recovers the Lagrange-d'Alembert equations with time-dependent nonlinear constraints as given in \eqref{evo_eqn_nh1} in absence of external forces. In particular, we notice that the second equation in \eqref{local_Dirac_Eqn_final} recovers the kinematic constraints $C_K$, although only $C_V$ was used to introduce the Dirac stricture $D_{\Delta_\mathcal{P}}$. This is due to the special link between the constraints $C_V$ and $C_K$ of thermodynamic type, see Definition \ref{def_thermotype}. We summarize the obtained results in the following theorem.

\begin{theorem}\label{DiracPont_theorem}  Given a variational constraint $C_{V} \subset  (\mathbb{R} \times T\mathcal{Q}) \times_{Y} TY$ as in \eqref{CV}, consider the induced Dirac structure $D_{\Delta_ \mathcal{P}}$ on $\mathcal{P}=(\mathbb{R} \times T\mathcal{Q}) \times_{Y} T^{\ast}Y $ as in \eqref{Dirac_P}. Let $L:\mathbb{R}  \times T\mathcal{Q} \rightarrow \mathbb{R}  $ be a time-dependent Lagrangian and $\mathcal{E} : \mathcal{P} \rightarrow \mathbb{R}  $ be the associated covariant generalized energy. Then the following statements are equivalent:
\begin{itemize}
\item The curve $\mathrm{x}(t)=(t,x(t),v(t),\mathsf{p}(t), p(t)) \in \mathcal{P}$ satisfies
\[
\begin{aligned}
&\dot{x}^i=v^i, \qquad p_i=\frac{\partial L}{\partial v^i}(t,x,v),\qquad \dot{p}_i-\frac{\partial L}{\partial x^i}(t,x,v)=\lambda_{r} A^{r}(t,x,v),\\
&  A_{i}^{r}(t,x,v)\dot{x}^{i} +B^{r}(t,x,v)=0,\qquad\dot{\mathsf{p}}-\frac{\partial L}{\partial t} =\lambda_{r}B^{r}(t,x,v).
\end{aligned}
\]
\item The curve $\mathrm{x}(t)=(t,x(t),v(t),\mathsf{p}(t), p(t)) \in \mathcal{P}$ satisfies the time-dependent Dirac system
\begin{equation}\label{DiracCond_P}
\big(\dot{\mathrm{x}}(t), \mathbf{d}\mathcal{E}(\mathrm{x}(t))\big) \in D_{\Delta_{\mathcal{P}}}(\mathrm{x}(t)).
\end{equation}
\end{itemize}
In other words, the Dirac system implies the system of equations \eqref{local_Dirac_Eqn_final} for the curve $x(t)\in\mathcal{Q}$. In particular, the curve $x(t)$ satisfies the kinematic constraint $C_K$.
\end{theorem} 

Let us recall that the Lagrange-d'Alembert equations with time-dependent and nonlinear nonholonomic constraints given in \eqref{local_Dirac_Eqn_final} are the general abstract type of equations that govern the time evolution of open simple thermodynamic systems, as explained in \S\ref{subsec_gen_setting}. Later in \S\ref{Section_4}, we will apply this Dirac formulation to open thermodynamic systems.

\paragraph{Energy balance equations.} From the equations \eqref{local_Dirac_DS} we deduce that the covariant generalized energy $\mathcal{E}(t,x,v ,\mathsf{p},p)$ defined in \eqref{CovariantEnergy} is preserved along the solution curve $\mathrm{x}(t)=(t,x(t),v(t),\mathsf{p}(t), p(t)) $ of the Dirac dynamical system \eqref{local_Dirac_DS},
\begin{equation}\label{conservation_cov_E}
\frac{d}{dt}\mathcal{E}(t,x,v,\mathsf{p}, p)=0.
\end{equation}
Note that $\mathcal{E}$ does not represent the total energy of the system. In fact, the total energy is represented by the generalized energy $E$ in \eqref{gen_energy}. In terms of $E$, equation \eqref{conservation_cov_E} are to be
\begin{equation}\label{energy_balance_E}
\frac{d}{dt}E(t,x,v, p)= - \frac{d}{dt}\mathsf{p}= - \frac{\partial L}{\partial t}(t,x,v) -\lambda_{r}B^{r}(t,x,v).
\end{equation}
This is the \textit{balance of energy} for the Dirac system. Note that $\frac{d}{dt}\mathsf{p}$ is interpreted as the power flowing out of the system. The first term on the right hand side is uniquely due to the explicit dependence of the Lagrangian on time. The second term is due to the affine characteristic of the kinematic constraint and will be interpreted later as the energy flowing in or out of the systems though its ports in the context of open systems. For the case where there does not exist any constraint, the energy balance equation of time-dependent mechanics can be recovered (see \cite{La1970}).

It is interesting to note that the equation for $\mathsf{p}$ is solved apart from the other equations. A natural initial condition for $\mathsf{p}$ is $\mathsf{p}(0)= - E(0)$, so that covariant generalized energy vanishes, i.e., $\mathcal{E}(t,x,v,\mathsf{p}, p)=0$ for all $t$, which is the {\it generalized energy analogue of the super-Hamiltonian constraint}.

\begin{remark}[Time-dependent Dirac systems]{\rm Let us stress that \eqref{DiracDyn_P} is called a time-dependent Dirac system, since it explicitly includes the time $t$ as a variable. The time-dependent Dirac structure $D_{\Delta _\mathcal{P}}\subset T\mathcal{P}\oplus T^*\mathcal{P}$ is defined at each point $\mathrm{x}=(t,x,v,\mathsf{p},p)\in\mathcal{P}$ of the covariant Pontryagin bundle, which includes the time $t$ as a variable.
This is in contrast with the Dirac structure appearing in mechanics $D_{\Delta _P}\subset TP\oplus T^*P$, which are defined at each point $(x,v,p)\in P$ of the Pontryagin bundle $P=T\mathcal{Q}\,\oplus\, T^*\mathcal{Q}$ and cannot incorporate time-dependent constraints.

From the field theoretic point of view, the time-dependent Dirac system \eqref{DiracDyn_P} can be interpreted as a special instance of a \textit{multi-Dirac formulation} for constrained field theories that extend the multi-Dirac field theory developed in \cite{VaYoLe2012}.
}
\end{remark}

\paragraph{A first Lagrange-d'Alembert-Pontryagin principle.} A first version of this principle can be obtained from \eqref{Vcond_abstract}--\eqref{VC_abstract} by a direct extension of the Lagrange-d'Alembert-Pontryagin principle for mechanical systems with linear constraints recalled in \eqref{LdAP_linear}. This variational principle, which uses the generalized energy $E(t,x,v,p)$ on $\mathbb{R}\times (T\mathcal{Q}\oplus T^*\mathcal{Q})$, is given as
\begin{equation}\label{LdAP}
\delta \int_{t_{1}}^{t_{2}} \Big[ \big\langle p, \dot x\big\rangle  -E(t,x,v,p)\Big]dt=0,
\end{equation}
subject to the kinematic constraints
\begin{equation}\label{KConstr_LdA}
A_{i}^{r}(t,x,v)\dot{x}^{i} +B^{r}(t,x,v)=0,\;\; r=1,...,m
\end{equation}
and for variations $\delta x$, $\delta v$, $\delta p$ subject to the variational constraints
\begin{equation}\label{VConstr_LdA}
A_{i}^{r}(t,x,v)\delta x^ i=0,\;\; r=1,...,m
\end{equation}
with $\delta x (t_1)=\delta x (t_ 2)=0$.

The principle in \eqref{LdAP} is defined on curves $t\mapsto (x(t),v(t), p(t))\in T\mathcal{Q}\oplus T^*\mathcal{Q}$.
From the stationarity conditions, it follows
\begin{equation}\label{LdAP_conditions}
\delta{p}:\;\;\dot x^i=v^i, \qquad \;\delta{x} : \;\;\dot p_i- \frac{\partial L}{\partial x^i}=\lambda_{r}A^{r}(t,x,v), \qquad \;\;\delta{v}:\; p_i= \frac{\partial L}{\partial v^i}.
\end{equation}

One notes that only a subset of the conditions associated to the Dirac dynamical system \eqref{local_Dirac_DS} are recovered, namely, the equation for $\mathsf{p}$ is missing.
In order to include this equation as a stationarity condition, we shall first express the Lagrange-d'Alembert-Pontryagin principle \eqref{LdAP} in the intrinsic form, as in \eqref{LdAP_theta}, by using the one form $\Theta_\mathcal{P}$ on $\mathcal{P}$ naturally induced from the canonical one-form $\Theta_{T^*Y}$ on $T^*Y$, namely,
\[
\Theta_{\mathcal{P}}=\pi_{(\mathcal{P}, T^{\ast}Y)}^{\ast}\Theta_{T^{\ast}Y},
\]
locally given by $\Theta_{\mathcal{P}}=p_{i}dx^{i}+\mathsf{p}dt$. Using $\Theta_\mathcal{P}$ and the covariant generalized energy $\mathcal{E}$ in \eqref{CovariantEnergy} the variational principle \eqref{LdAP}--\eqref{VConstr_LdA} can be intrinsically written as follows for sections of the form $\mathrm{x}(t)=(t, x(t), v(t), \mathsf{p}(t), p(t))$:
\begin{equation}\label{intrinsic_LDAP}
\delta \int_{t_{1}}^{t_{2}} \Big[ \big\langle\Theta_{\mathcal{P}}(\mathrm{x}(t)), \dot{\mathrm{x}}(t)\big\rangle-\mathcal{E}(\mathrm{x}(t))\Big]dt=0,
\end{equation}
subject to the kinematic and variational constraints
\begin{equation}\label{intrinsic_CV_CK}
\dot{\mathrm{x}}(t) \in \Delta_{\mathcal{P}}(\mathrm{x}(t))\quad\text{and}\quad 
\delta{\mathrm{x}}(t) \in \Delta_{\mathcal{P}}(\mathrm{x}(t))
\end{equation}
with the endpoint conditions $T\pi_{(\mathcal{P},Y)}(\delta\mathrm{x}(t_{1}))=T\pi_{(\mathcal{P},Y)}(\delta\mathrm{x}(t_{2}))=0 $.

One indeed notes the equalities
\begin{equation}\label{cancellation}
\begin{aligned}
\big\langle\Theta_{\mathcal{P}}(\mathrm{x}), \dot{\mathrm{x}}\big\rangle-\mathcal{E}(\mathrm{x})&= \big\langle p, \dot x\big\rangle  + \mathsf{p} - (\mathsf{p}+E(t,x,v,p))\\
&= \big\langle p, \dot x\big\rangle  -E(t,x,v,p)=\big\langle p, \dot x-v\big\rangle  +L(t,x,v)
\end{aligned}
\end{equation}
and, from \eqref{delta_P}, the equivalences
\[
\dot{\mathrm{x}}(t) \in \Delta_{\mathcal{P}}(\mathrm{x}(t))\;\;\Leftrightarrow\;\;\eqref{KConstr_LdA}\quad\quad\text{and}\quad\quad \delta{\mathrm{x}}(t) \in \Delta_{\mathcal{P}}(\mathrm{x}(t))\;\;\Leftrightarrow\;\;\eqref{VConstr_LdA}
\]
that hold since $\dot{\mathrm{x}}(t)=(1, \dot x(t), \dot v(t), \dot{\mathsf{p}}(t), \dot p(t))$ and $\delta \mathrm{x}(t)= (0, \delta x(t), \delta v(t), \delta \mathsf{p}(t), \delta p(t))$. As before, the stationary conditions for \eqref{intrinsic_LDAP} are given by \eqref{LdAP_conditions} and do not recover the equation for $\mathsf{p}$ in \eqref{local_Dirac_DS}.
This is due to the fact that the curve $\mathrm{x}(t)$ \textit{and} its variations are sections, see Remark \ref{curve_section_remark}, namely, we have
\begin{equation}\label{delta_x_vertical}
\begin{aligned}
\delta\mathrm{x}(t)&=\left.\frac{d}{d\varepsilon}\right|_{\varepsilon=0} \mathrm{x}_\varepsilon(t)= \left.\frac{d}{d\varepsilon}\right|_{\varepsilon=0} \big(t, x_\varepsilon(t), v_\varepsilon(t),\mathsf{p}_\varepsilon(t), p_\varepsilon(t)\big)\\
&=(0, \delta x(t), \delta v(t), \delta \mathsf{p}(t), \delta p(t))
\end{aligned}
\end{equation}
and thus the first component of $\delta \mathrm{x}(t)$ must be zero. In the language of fiber bundles, we say that $\delta \mathrm{x}(t)$ is vertical, relative to the projection $\mathcal{P}\rightarrow \mathbb{R}$.

\medskip

\paragraph{The Lagrange-d'Alembert-Pontryagin principle associated to the Dirac system.} In order to get all the equations for the Dirac system in \eqref{local_Dirac_DS} from the variational formulation, we shall define the action functional \eqref{intrinsic_LDAP} on arbitrary curves $\mathrm{x}(\tau)$ in the covariant Pontryagin bundle $\mathcal{P}$, namely,
\begin{equation}\label{y_explicit}
\mathrm{x}(\tau)= (t(\tau), x(\tau), v(\tau), \mathsf{p}(\tau), p(\tau))\in \mathcal{P},
\end{equation}
rather than just on sections, while we still require that the critical curve is a section, i.e. $t(\tau)=\tau$. We shall denote by $\mathrm{x}'$ the derivative with respect to $\tau$. For such curves, we consider the same Lagrange-d'Alembert-Pontryagin principle as in \eqref{intrinsic_LDAP}--\eqref{intrinsic_CV_CK}, and seek for a \textit{section} $\mathrm{x}(t)=(t, x(t),v(t), \mathsf{p}(t), p(t))$ critical for
\begin{equation}\label{intrinsic_LDAP_tau}
\delta \int_{\tau_{1}}^{\tau_{2}} \Big[ \big\langle\Theta_{\mathcal{P}}(\mathrm{x}(\tau)), \mathrm{x}'(\tau)\big\rangle-\mathcal{E}(\mathrm{x}(\tau))\Big]d\tau=0,
\end{equation}
subject to the kinematic and variational constraints
\begin{equation}\label{CV_CK_general}
\dot{\mathrm{x}}(t) \in \Delta_{\mathcal{P}}(\mathrm{x}(t))\quad\text{and}\quad 
\delta{\mathrm{x}}(t) \in \Delta_{\mathcal{P}}(\mathrm{x}(t)),
\end{equation}
with the endpoint conditions $T\pi_{(\mathcal{P},Y)}(\delta\mathrm{x}(\tau_{1}))=T\pi_{(\mathcal{P},Y)}(\delta\mathrm{x}(\tau_{2}))=0 $.

Equivalently, the critical point condition \eqref{intrinsic_LDAP_tau} reads
\[
\left.\frac{d}{d\varepsilon}\right|_{\varepsilon=0}\int_{\tau_{1}}^{\tau_{2}} \Big[ \big\langle\Theta_{\mathcal{P}}(\mathrm{x}_\varepsilon(\tau)), \mathrm{x}'_\varepsilon(\tau)\big\rangle-\mathcal{E}(\mathrm{x}_\varepsilon(\tau))\Big]d\tau=0,
\] 
where $\mathrm{x}_\varepsilon(\tau)=( t_\varepsilon(\tau), x_\varepsilon(\tau),v_\varepsilon(\tau), \mathsf{p}_\varepsilon(\tau), p_\varepsilon(\tau))$ is such that $t_{\varepsilon=0}(\tau)=\tau$, for all $\tau$.
Taking arbitrary variations
\[
\delta\mathrm{x}(t)=\left.\frac{d}{d\varepsilon}\right|_{\varepsilon=0} \mathrm{x}_\varepsilon(t)=(\delta t(t), \delta x(t), \delta v(t), \delta \mathsf{p}(t), \delta p(t))
\]
which are not necessarily vertical, (compare to \eqref{delta_x_vertical}), this variational formulation yields the equations
\begin{equation}\label{intrinsic_LDAPEqn}
\mathbf{i}_{\dot{\mathrm{x}}}\Omega_{\mathcal{P}}-\mathbf{d}\mathcal{E}(\mathrm{x}) \in 
\Delta_{\mathcal{P}}(\mathrm{x})^{\circ},\qquad \dot{\mathrm{x}}\in \Delta_{\mathcal{P}}(\mathrm{x}).
\end{equation}
These equations are equivalent to the condition of the Dirac system, namely, $\big(\dot{\mathrm{x}}, \mathbf{d}\mathcal{E}(\mathrm{x})\big) \in D_{\Delta_{\mathcal{P}}}(\mathrm{x})$.
\medskip

We now write explicitly this variational condition. As opposed to \eqref{cancellation}, we have
\[
\big\langle\Theta_{\mathcal{P}}(\mathrm{x}), \mathrm{x}'\big\rangle-\mathcal{E}(\mathrm{x})= \big\langle p,  x'\big\rangle  + \mathsf{p} t'- \mathcal{E}(t,x,v,\mathsf{p},p)= \big\langle p,  x'-v\big\rangle  + \mathsf{p}(t'-1)-L(t,x,v).
\]
The Lagrange-d'Alembert-Pontryagin principle in \eqref{intrinsic_LDAP_tau}--\eqref{CV_CK_general} for a curve $\mathrm{x}(\tau)$ given in \eqref{y_explicit} reads explicitly
\begin{equation}\label{LDAP_TNS}
\delta \int_{\tau_{1}}^{\tau_{2}} \Big[\big\langle p, x^{\prime} \big\rangle +\mathsf{p} t^{\prime} -\mathcal{E}\big(t,x,v,\mathsf{p},p\big)\Big]d\tau=0,
\end{equation}
subject to the kinematic constraints
\begin{equation}\label{CK_TNS}
A_{i}^{r}\big(t ,x ,v \big)\dot x^{i} +B^{r}\big(t ,x ,v \big) =0, \;\; r=1,...,m,
\end{equation}
for variations subject to the variational constraints
\begin{equation}\label{CV_TNS}
A_{i}^{r}\big(t ,x ,v \big)\delta{x} ^{i}+B^{r}\big(t ,x ,v  \big)\delta{t} =0, \;\; r=1,...,m.
\end{equation}

Note that \eqref{CK_TNS} is imposed on the critical curve, which is a section, this is why $x'=\dot x$ and $t'=\dot t=1$. On general curves this constraint would read $A_{i}^{r}\big(t ,x ,v \big)\dot x'{}^{i} +B^{r}\big(t ,x ,v \big)t' =0$.

A direct application of \eqref{LDAP_TNS}--\eqref{CV_TNS} yields the following equations
\begin{equation}\label{vari_time_dependent}
\begin{aligned}
&\delta{p}:\;\; x'=v, \qquad \delta{\mathsf{p}}:\;\; t'=1, \qquad \delta{x}: \;p'- \frac{\partial L}{\partial x}=\lambda_{r}A^{r}(t,x,v), \\
&\delta{v}:\; \;p= \frac{\partial L}{\partial v}, \qquad \delta{t}:\;\;\mathsf{p}'=\frac{\partial L}{\partial t}+\lambda_{r}B^{r}(t,x,v),
\end{aligned}
\end{equation}
together with $A_{i}^{r}(t,x,v)(x^i)' +B^{r}(t,x,v)t'=0$. These equations are the local expressions of \eqref{intrinsic_LDAPEqn}.
\medskip

From the equations \eqref{intrinsic_LDAPEqn}, we also get the equivalence between the Lagrange-d'Alembert-Pontryagin principle in \eqref{intrinsic_LDAP_tau}--\eqref{CV_CK_general} and the Dirac dynamical system in \eqref{DiracDyn_P} written for arbitrary curves $\mathrm{x}(\tau)\in \mathcal{P}$. This is the statement of the next theorem.

\begin{theorem}[Equivalence of the Dirac and variational formulation]\label{Dirac_Variational_theorem}  
The following statements on a section $\mathrm{x}(t)\in \mathcal{P}$ are equivalent:
\begin{itemize}
\item The section $\mathrm{x}(t)$ is a solution of the Dirac dynamical system
\[
\big(\dot{\mathrm{x}}, \mathbf{d}\mathcal{E}(\mathrm{x})\big) \in D_{\Delta_{\mathcal{P}}}(\mathrm{x})
\] 
\item The section $\mathrm{x}(t)$ satisfies
\[
\mathbf{i}_{\dot{\mathrm{x}}}\Omega_{\mathcal{P}}-\mathbf{d}\mathcal{E}(\mathrm{x}) \in 
\Delta_{\mathcal{P}}(\mathrm{x})^{\circ},\qquad \dot{\mathrm{x}} \in \Delta_{\mathcal{P}}(\mathrm{x})
\]
\item The section $\mathrm{x}(t)$ is a critical point of  the variational formulation \eqref{intrinsic_LDAP_tau}--\eqref{CV_CK_general}.
\end{itemize}
Moreover, the Dirac dynamical system deduces the system of equations for a curve $x(t) \in \mathcal{Q}$ as in \eqref{local_Dirac_Eqn_final}.
\end{theorem} 

\begin{remark}[Inclusion of external forces]{\rm External forces $\mathcal{F}^{\rm ext}: \mathbb{R}\times T \mathcal{Q}\rightarrow T^*\mathcal{Q}$ can be included in all the formulations in \S\ref{TDCPB}, consistently with the equations obtained from \eqref{Vcond_abstract} as follows.

Suppose that an external force field $\mathcal{F}^{\rm ext}: \mathbb{R}\times T \mathcal{Q}\rightarrow T^*\mathcal{Q}$, with $\mathcal{F}^{\rm ext}(t,x,v)\in T^*_x\mathcal{Q}$, for all $(t,x,v)\in\mathbb{R}\times T\mathcal{Q}$ is given. Consider the natural projection $\pi_{(\mathcal{P}, \mathcal{Q})}: \mathcal{P} \to \mathcal{Q}$ as $(t,x,v, \mathsf{p}, p) \mapsto x$, 
the external force field  $\mathcal{F}^{\rm ext}$ on $T^\ast\mathcal{Q}$ can be lifted as a horizontal one-form on $\mathcal{P}$ as
$$
\widetilde{\mathcal{F}}^{\rm ext}(t,x,v, \mathsf{p}, p) \cdot W=\left<\mathcal{F}^{\rm ext}(t, x,v), T_{(t,x,v, \mathsf{p}, p)}\pi_{(\mathcal{P}, \mathcal{Q})}(W)\right>,
$$
where $W \in T_{(t,x,v, \mathsf{p}, p)}\mathcal{P}$. Locally, we have $\widetilde{\mathcal{F}}^{\rm ext}(t,x,v, \mathsf{p}, p)=(t,x,v, \mathsf{p}, p,0,\mathcal{F}^{\rm ext}(t,x,v),0,0,0)$. 

Therefore, the Dirac dynamical system in \eqref{DiracCond_P} may be replaced by 
\begin{equation}\label{DiracCond_P_ext}
\big(\dot{\mathrm{x}}(t), \mathbf{d}\mathcal{E}(\mathrm{x}(t))-\widetilde{\mathcal{F}}^{\rm ext}(\mathrm{x}(t))\big) \in D_{\Delta_{\mathcal{P}}}(\mathrm{x}(t)).
\end{equation}

The associated variational formulation is given by the Lagrange-d'Alembert-Pontryagin principle for arbitrary curves $\mathrm{x}(\tau)= (t(\tau), x(\tau), v(\tau), \mathsf{p}(\tau), p(\tau))$ in the covariant Pontryagin bundle $\mathcal{P}$ seeking for a \textit{section} $\mathrm{x}(t)=(t, x(t),v(t), \mathsf{p}(t), p(t))$, which is critical for
\begin{equation}\label{intrinsic_LDAP_tau_ext}
\delta \int_{\tau_{1}}^{\tau_{2}} \Big[ \big\langle\Theta_{\mathcal{P}}(\mathrm{x}(\tau)), \mathrm{x}'(\tau)\big\rangle-\mathcal{E}(\mathrm{x}(\tau))\Big]d\tau+
\int_{\tau_{1}}^{\tau_{2}} \widetilde{\mathcal{F}}^{\rm ext}(\mathrm{x}(\tau)) \cdot \delta{\mathrm{x}}(\tau) d\tau=0,
\end{equation}
subject to the kinematic and variational constraints in \eqref{CV_CK_general} with the endpoint conditions $T\pi_{(\mathcal{P},Y)}(\delta\mathrm{x}(\tau_{1}))=T\pi_{(\mathcal{P},Y)}(\delta\mathrm{x}(\tau_{2}))=0 $, where $\tau(t)=t$ and $\mathrm{x}'$ the derivative with respect to $\tau$. 

Thus, the Dirac formulation in \eqref{DiracCond_P_ext} and the variational formulation in \eqref{intrinsic_LDAP_tau_ext} provide the same evolution equations
\[
\mathbf{i}_{\dot{\mathrm{x}}}\Omega_{\mathcal{P}}-\mathbf{d}\mathcal{E}(\mathrm{x}) + \widetilde{\mathcal{F}}^{\rm ext}(\mathrm{x})\in 
\Delta_{\mathcal{P}}(\mathrm{x})^{\circ},\qquad \dot{\mathrm{x}} \in \Delta_{\mathcal{P}}(\mathrm{x}),
\]
which are given in coordinates by
\[
\begin{aligned}
&\dot{x}^i=v^i, \qquad p_i=\frac{\partial L}{\partial v^i}(t,x,v),\qquad \dot{p}_i-\frac{\partial L}{\partial x^i}(t,x,v)=\lambda_{r} A^{r}(t,x,v)+\mathcal{F}^{\rm ext}(t, x,v),\\
&  A_{i}^{r}(t,x,v)\dot{x}^{i} +B^{r}(t,x,v)=0,\qquad\dot{\mathsf{p}}-\frac{\partial L}{\partial t} =\lambda_{r}B^{r}(t,x,v).
\end{aligned}
\]

}
\end{remark}

\subsection{Lagrange-Dirac systems on the cotangent bundle}
 
\begin{figure}[h]
\begin{center}
\includegraphics[scale=.53]{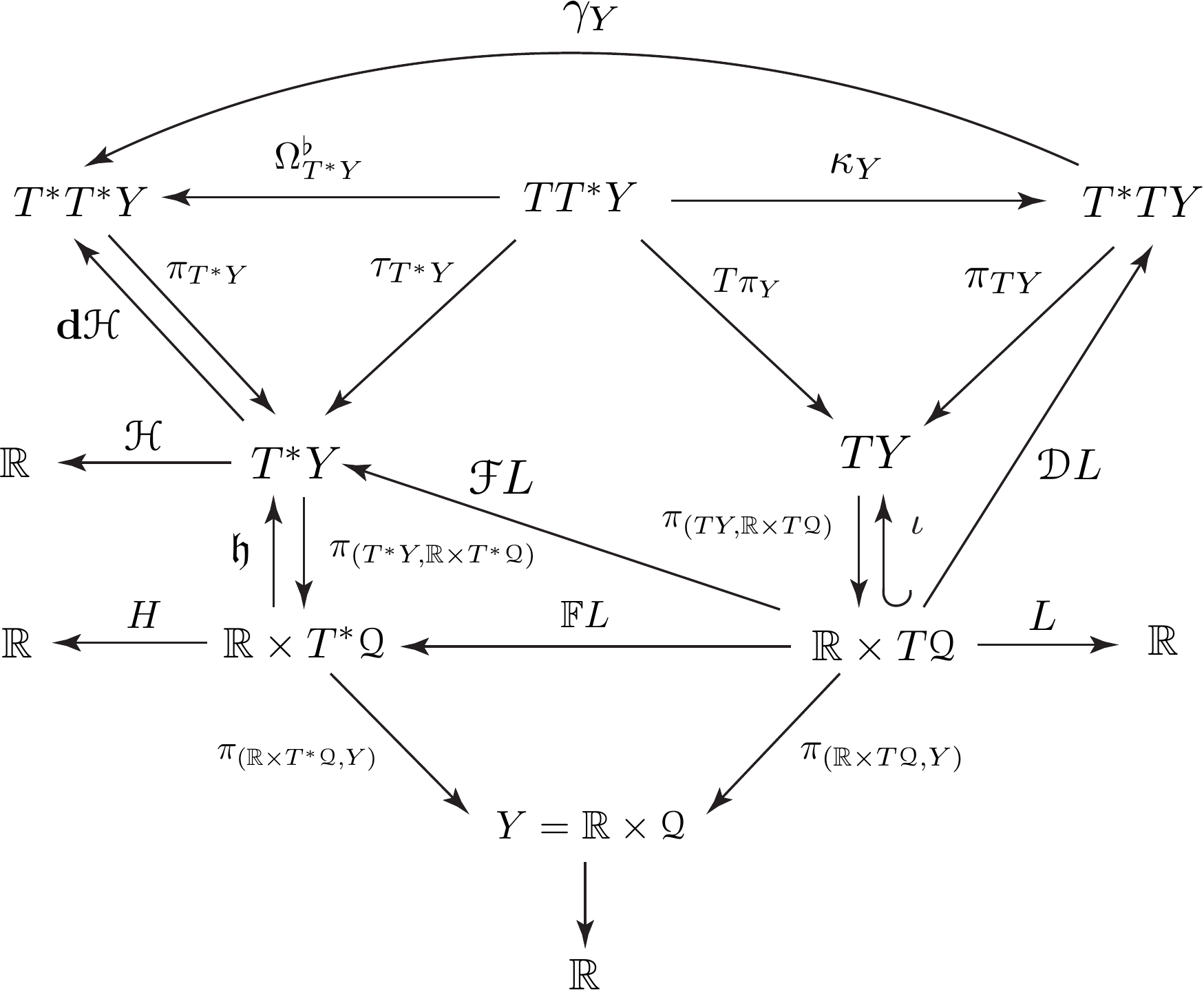}
\caption{A diagram illustrating the canonical diffeomorphisms and the bundle projections}
\label{BundPic}
\end{center}
\end{figure}

As in mechanics, we can construct another type of Dirac dynamical systems, called {\it Lagrange-Dirac systems}, based on an induced Dirac structure on the cotangent bundle $T^{\ast}Y$ as well as the Lagrangian $L$ (rather than the covariant generalized energy $\mathcal{E}$). In fact, given a time-dependent Lagrangian $L:\mathbb{R}\times T\mathcal{Q}\rightarrow\mathbb{R}$, a key step of this formulation is the construction of the map $\mathcal{D}L: \mathbb{R} \times T\mathcal{Q} \to T^\ast TY$, which allows to define a Dirac differential $\mathbf{d}_DL: \mathbb{R} \times T\mathcal{Q} \to T^\ast T^\ast Y$ for the time-dependent Lagrangian. To do this, we shall use the iterated tangent and cotangent bundles over the extended configuration manifold $Y=\mathbb{R} \times \mathcal{Q}$, as illustrated in Fig. \ref{BundPic}.

\paragraph{Induced Dirac structure on the cotangent bundle $T^{\ast}Y$.} Now we shall define the Dirac structure on $T^\ast Y$ induced from a given variational constraint $C_{V} \subset (\mathbb{R} \times T\mathcal{Q}) \times_{Y} T(\mathbb{R}\times \mathcal{Q})$ as given in \eqref{CV}.

First, consider the variational constraint $\mathscr{C}_{V} \subset (\mathbb{R} \times T^{\ast}\mathcal{Q}) \times_{Y} TY$ defined from $C_V$, for each $(t,x,p)\in \mathbb{R} \times T^{\ast}\mathcal{Q}$, by
\begin{equation}\label{scrCV}
\mathscr{C}_{V}(t,x,p):=C_{V}(t,x,v) \subset T_{(t,x)}Y,
\end{equation}
where $v$ is determined such that $\frac{\partial L}{\partial v}(t,x,v)=p$. At first glance, it appears that the definition of $\mathscr{C}_V$ from $C_V$ is only possible when the Lagrangian is nondegenerate. In fact, it only needs the nondegeneracy with respect to the velocity $v$ that appears explicitly in $C_V$. As will be shown in \S\ref{Section_4}, the definition of $\mathscr{C}_V$ is always possible in thermodynamics, due to the specific form of the Lagrangian, even though it is degenerate.

Let $\pi_{Y}: T^{\ast}Y \to Y$, $(t,x,\mathsf{p}, p) \mapsto (t,x)$ be the cotangent bundle projection. The constraint distribution $\Delta_{T^{\ast}Y}$ on $T^{\ast}Y$ is defined by
\begin{equation}\label{def_D_CotY}
\Delta_{T^{\ast}Y}(t, x, \mathsf{p}, p)=\left(T_{(t, x, \mathsf{p}, p)}\pi_{Y}\right)^{-1}\big(\mathscr{C}_{V}(t, x, \mathsf{p}, p)\big) \subset T_{(t, x, \mathsf{p}, p)}T^{\ast}Y,
\end{equation}
for each $(t, x, \mathsf{p}, p)\in T^*Y$. If $C_V$ is given as in \eqref{CV}, then it follows from \eqref{scrCV} and \eqref{def_D_CotY} that the distribution reads
\begin{align*}
\Delta_{T^{\ast}Y}(t, x, \mathsf{p}, p)&=\big\{(\delta{t},\delta{x},\delta{\mathsf{p}}, \delta{p}) \in T_{(t, x, \mathsf{p}, p)}T^{\ast}Y \mid \nonumber \\ 
& \hspace{2cm} \mathcal{A}_{i}^{r}(t,x,p)\delta{x}^{i} +\mathcal{B}(t,x,p)\delta{t}=0,\; r=1,...,m\big\}.
\end{align*} 

Further, from the distribution $\Delta_{T^{\ast}Y}$ and the canonical symplectic form $\Omega_{T^\ast Y}$, the induced Dirac structure $D_{\Delta_{T^{\ast}Y}}$ on $T^*Y$ is defined as in \eqref{def_Dirac} by
\begin{equation}\label{Dirac_CotY}
\begin{aligned}
D_{\Delta_{T^{\ast}Y}}(\mathrm{z})&=\big\{(\mathfrak{u}_{\mathrm{z}}, \mathfrak{a}_{\mathrm{z}}) \in T_{\mathrm{z}}T^{\ast}Y \times T_{\mathrm{z}}^{\ast}T^{\ast}Y \mid \mathfrak{u}_{\mathrm{z}} \in \Delta_{T^{\ast}Y}(\mathrm{z}),  \\
& \hspace{3cm} \left<\mathfrak{a}_{\mathrm{z}}, \mathfrak{v}_{\mathrm{z}}\right>= \Omega_{T^{\ast}Y}(\mathrm{z})(\mathfrak{u}_{\mathrm{z}}, \mathfrak{v}_{\mathrm{z}}),\;\forall \; \mathfrak{v}_{\mathrm{z}} \in \Delta_{T^{\ast}Y}(\mathrm{z})\big\}.
\end{aligned}
\end{equation}
For $\mathrm{z}=(t,x,\mathsf{p}, p) \in T^{\ast}Y$, we write $\mathfrak{u}_{\mathrm{z}}=( u_t, u_x, u_{\mathsf{p}}, u_p)\in T_{\mathrm{z}}T^{\ast}Y$, $\mathfrak{v}_{\mathrm{z}}=( \delta{t}, \delta{x},  \delta{\mathsf{p}}, \delta{p}) \in T_{\mathrm{z}}T^{\ast}Y$ and $\mathfrak{a}_{\mathrm{z}}=(\pi, \alpha, \gamma, w) \in T^{\ast}_{\mathrm{z}}T^{\ast}Y$.
Then, the condition $(\mathfrak{u}_{\mathrm{z}}, \mathfrak{a}_{\mathrm{z}}) \in D_{\Delta_{T^{\ast}Y}}(\mathrm{z})$ reads
\begin{equation}\label{DiracCond_CotY}
w=u_x, \quad\gamma=u_t,\quad (t,x,u_t,u_x) \in \mathscr{C}_{V}(t, x, p),\quad(u_{\mathsf{p}}+\pi, u_p+\alpha) \in \mathscr{C}_{V}(t, x, p)^{\circ}.
\end{equation}
In local coordinates, using the Lagrange multipliers $\lambda_{r},\; r=1,...,m$, we get
\begin{equation}\label{local_Dirac_P_new}
\begin{split}
&u_x^i=w^i, \qquad u_t=\gamma,\qquad  \mathcal{A}_{i}^{r}(t,x,p)u_x^{i} +\mathcal{B}^{r}(t,x,p)u_t=0,\\[3mm]
& u_{\mathsf{p}}+\pi =\lambda_{r}\mathcal{B}^{r}(t,x,p),\qquad (u_p)_i+\alpha_i=\lambda_{r} \mathcal{A}^{r}_i(t,x,p).
\end{split}
\end{equation}

\paragraph{The covariant Legendre transform.} Given a time-dependent Lagrangian $L:\mathbb{R}  \times T\mathcal{Q}\rightarrow\mathbb{R}$, possibly degenerate, the {\it covariant Legendre transform} $\mathcal{F}L: \mathbb{R} \times T\mathcal{Q} \to  T^{\ast}Y$ is defined by  
\begin{equation}\label{covLegendreTrans}
\mathcal{F}L(t,x,v)=\left(t,x, L-\left<\frac{\partial L}{\partial v}, v\right>, \frac{\partial L}{\partial v}\right).
\end{equation}
This follows the general expression of the covariant Legendre transform used in field theories, \cite{GIMM1997}.
The corresponding element $(t,x, \mathsf{p}, p)$ in $T^{\ast}Y$ is thus given by
\[
\mathsf{p}= L-\left<\frac{\partial L}{\partial v}, v\right>=-E_L\quad\text{and}\quad p=\frac{\partial L}{\partial v},
\]
where $E_L$ is the Lagrangian energy, see \eqref{E_L}.

\paragraph{The iterated tangent and cotangent bundles over $Y=\mathbb{R} \times \mathcal{Q}$.} Recall from \cite{YoMa2006a} that there exists three symplectomorphisms among the iterated tangent and cotangent bundles $TT^{\ast}Y$, $T^{\ast}TY$, and $T^{\ast}T^{\ast}Y$, which were originally considered by \cite{Tu1977} in the context of the generalized Legendre transform.

Let $\tau_Y: TY \to Y$ denote the tangent bundle projection, locally given by $(t,x, \delta{t},\delta{x}) \mapsto (t,x)$. 
We have $\tau_Y=\pi_{(\mathbb{R}\times T\mathcal{Q}, Y)} \circ \pi_{(TY, \mathbb{R} \times T\mathcal{Q})}$, for the projections $\pi_{(TY, \mathbb{R} \times T\mathcal{Q})}: TY \to \mathbb{R} \times T\mathcal{Q}$, $(t,x, \delta{t},\delta{x}) \mapsto (t,x,\delta{x})$ and $\pi_{(\mathbb{R}\times T\mathcal{Q}, Y)}:\mathbb{R}\times T\mathcal{Q}\rightarrow  Y$, $\pi_{(\mathbb{R}\times T\mathcal{Q}, Y)}(t,x,\delta x)=(t,x)$.
Let $\pi_{Y}: T^{\ast}Y \to Y$ denote the cotangent bundle projection, locally given by $(t,x,\mathsf{p}, p) \mapsto (t,x)$. Similarly as before, we have $\pi_Y=\pi_{(\mathbb{R}\times T^\ast \mathcal{Q}, Y)} \circ \pi_{(T^\ast Y, \mathbb{R} \times T^\ast \mathcal{Q})}$.

Denoting by $(t,x,\mathsf{p}, p,\delta{t},\delta{x}, \delta{\mathsf{p}}, \delta{p})$ the local coordinates on $TT^{\ast}Y$, we have the canonical symplectomorphisms
\begin{align*}
\kappa_{Y}&: TT^{\ast}Y \to T^{\ast}TY, \;\; (t,x,\mathsf{p}, p,\delta{t},\delta{x}, \delta{\mathsf{p}}, \delta{p}) \mapsto (t,x, \delta{t},\delta{x}, \delta\mathsf{p}, \delta{p}, \mathsf{p},p), \\
\Omega^{\flat}_{T^{\ast}Y}&: TT^{\ast}Y \to T^{\ast}T^{\ast}Y, \;\; (t,x,\mathsf{p}, p,\delta{t},\delta{x}, \delta{\mathsf{p}}, \delta{p}) \mapsto (t,x,\mathsf{p}, p,-\delta{\mathsf{p}}, -\delta{p},\delta{t},\delta{x}),
\end{align*} 
as illustrated in Figure \ref{BundPic}, with
\[
\Omega_{TT^{\ast}Y}=(\Omega^{\flat}_{T^\ast Y})^\ast\Omega_{T^{\ast}T^{\ast}Y}=(\kappa_Y)^{\ast}\Omega_{T^{\ast}TY}. 
\]
Here $\Omega_{T^{\ast}T^{\ast}Y}$ and $\Omega_{T^{\ast}TY}$ are the canonical symplectic forms on $T^*T^*Y$ and $T^*TY$, while the symplectic form on $TT^*Y$ is given by $\Omega_{TT^{\ast}Y}=dq\wedge d\delta p+ d\delta q\wedge dp$.

Besides the cotangent bundle projections $\pi_{TY}: T^\ast TY \to TY, (t,x, \delta{t},\delta{x}, \delta\mathsf{p}, \delta{p}, \mathsf{p},p) \mapsto (t,x, \delta{t},\delta{x})$ and $\pi_{T^{\ast}Y}: T^{\ast}T^{\ast}Y \to T^{\ast}Y, (t,x,\mathsf{p}, p,-\delta{\mathsf{p}}, -\delta{p},\delta{t},\delta{x}) \mapsto (t,x,\mathsf{p}, p)$ and the  tangent bundle projection $\tau_{T^{\ast}Y}: TT^{\ast}Y \to T^{\ast}Y, (t,x,\mathsf{p}, p, \delta{t},\delta{x},\delta{\mathsf{p}}, \delta{p}) \mapsto (t,x,\mathsf{p}, p)$, we also need the tangent map of $\pi_Y$ given by 
$T\pi_{Y}: TT^{\ast}Y \to TY$, $(t,x,\mathsf{p}, p, \delta{t},\delta{x},\delta{\mathsf{p}}, \delta{p}) \mapsto (t,x, \delta{t}, \delta{x})$.

Further, we can consider the symplectomorphism $\gamma_Y: T^\ast TY \to T^\ast T^\ast Y$ defined by $\gamma_Y:= \Omega_{T^\ast Y}^\flat  \circ \kappa_Y^{-1}$, locally given by
\begin{equation}\label{local_gamma_Y}
\gamma_Y(t,x, \delta{t},\delta{x}, \delta\mathsf{p}, \delta{p}, \mathsf{p},p)= (t,x,\mathsf{p}, p,-\delta{\mathsf{p}}, -\delta{p},\delta{t},\delta{x}).
\end{equation}

\paragraph{The Dirac differential of the Lagrangian.} Given a time-dependent Lagrangian $L:\mathbb{R} \times T\mathcal{Q}\rightarrow \mathbb{R}$, we define a map $\mathcal{D}L: \mathbb{R} \times T\mathcal{Q} \to T^\ast TY$ by 
\begin{equation}\label{D_L}
\mathcal{D}L(t,x,v)=\left(t, x,1, v, \frac{\partial L}{\partial t}, \frac{\partial L}{\partial x}, -E_L, \frac{\partial L}{\partial v}\right),
\end{equation}
where $E_L(t,x,v)$ is the Lagrangian energy, given by $E_L(t,x,v)=\left<\frac{\partial L}{\partial v}, v \right> -L(t,x,v)$. Then, the {\it Dirac differential of the time-dependent Lagrangian} $L$, denoted $\mathbf{d}_DL: \mathbb{R} \times T\mathcal{Q} \to T^\ast T^\ast Y$, is defined by 
\[
\mathbf{d}_DL:=\gamma_Y \circ \mathcal{D}L,
\]
where $\gamma_Y: T^\ast TY \to T^\ast T^\ast Y$ is the symplectic diffeomorphism given in \eqref{local_gamma_Y}. Thus, using \eqref{local_gamma_Y} and \eqref{D_L}, we get the local expression of the map $\mathbf{d}_DL: \mathbb{R} \times T\mathcal{Q} \to T^\ast T^\ast Y$ as
\[
\mathbf{d}_DL(t,x,v)=\left(t, x, -E_L, \frac{\partial L}{\partial v}, -\frac{\partial L}{\partial t}, - \frac{\partial L}{\partial x},1, v \right).
\]
This is an extension of the Dirac differential for time-independent Lagrangians introduced in \cite{YoMa2006a}.

\paragraph{Lagrange-Dirac systems on $T^{\ast}Y$.} Given a variational constraint $C_V$, we define the induced distribution $\Delta_{T^*Y}$ on $T^*Y$ defined in \eqref{def_D_CotY} and the Dirac structure $D_{\Delta_{T^*Y}}$ as in \eqref{Dirac_CotY}. We consider a time-dependent Lagrangian $L:\mathbb{R} \times T\mathcal{Q}\rightarrow \mathbb{R}$.

\begin{definition}\label{def_Lagr_Dir}
Given $\Delta_{T^*Y}$ and $L$ as above, the associated {\bfi time-dependent Lagrange-Dirac dynamical system} for a curve
$\mathrm{x}(t)= (t,x(t),v(t), \mathsf{p}(t), p(t))\in \mathcal{P}$ is
\[
\left(\dot{\mathrm{z}}(t),\mathbf{d}_DL( \mathrm{y}(t))\right)\in 
D_{\Delta_{T^*Y}}(\mathrm{z}(t)),
\]
where $\mathrm{y}(t)=(t,x(t),v(t))\in \mathbb{R}\times T\mathcal{Q}$ and $\mathrm{z}(t)= (t,x(t),\mathsf{p}(t), p(t))\in T^*Y$.
\end{definition}
\medskip

More explicitly, the Lagrange-Dirac system reads
\begin{equation}\label{DiracSysCond_CotY}
\left((\dot{t}, \dot x(t), \dot{\mathsf{p}}(t), \dot p(t)),\mathbf{d}_DL( t, x(t), v(t))\right)\in 
D_{\Delta_{T^*Y}}(t,x(t),\mathsf{p}(t), p(t)).
\end{equation}
Using the local expression \eqref{local_Dirac_P_new} of the Dirac structure \eqref{Dirac_CotY} we get the equations of motion as
\begin{equation}\label{LD_equations}
\begin{aligned}
&\dot{x}^i=v^i, \quad  \dot{\mathsf{p}}-\frac{\partial L}{\partial t}(t,x,v) =\lambda_{r}\mathcal{B}^{r}(t,x,p),\quad \dot{p}_i-\frac{\partial L}{\partial x^i}(t,x,v)=\lambda_{r} \mathcal{A}^{r}(t,x,p),\\
&\mathcal{A}^{r}_{i}(t,x,p)\dot{x}^{i} +\mathcal{B}^{r}(t,x,p)=0,\quad \dot{t}=1, 
\end{aligned}
\end{equation}
together with the base point condition in \eqref{DiracSysCond_CotY} which gives the relations
\begin{equation}\label{BPC}
p_i=\frac{\partial L}{\partial v^i}\quad\text{and}\quad \mathsf{p}=- E_L= L- \frac{\partial L}{\partial v^i}v^i.
\end{equation}
Equations \eqref{LD_equations}-\eqref{BPC} yield the Lagrange-d'Alembert equations with time-dependent and nonlinear nonholonomic constraints considered in \eqref{evo_eqn_nh1}.

\medskip

The Lagrange-Dirac formulation of the Lagrange-d'Alembert equations \eqref{local_Dirac_Eqn_final} with nonlinear time-dependent constraint of thermodynamic type is summarized in the following.

\begin{theorem}\label{theorem_LD}  
 Consider a variational constraint $C_{V} \subset  (\mathbb{R} \times T\mathcal{Q}) \times_{Y} TY$ as in \eqref{CV} and the associated kinematic constraint $C_K$ of the thermodynamic type given in \eqref{CK}. Let  $L: \mathbb{R} \times T\mathcal{\mathcal{Q}}  \rightarrow \mathbb{R}  $ be a time-dependent Lagrangian, $ \mathscr{C} _V \subset    (\mathbb{R} \times T^*\mathcal{Q}) \times_{Y} TY$ be the variational constraint  as in \eqref{scrCV}, and define the induced Dirac structure $D_{\Delta_ {T^*Y }}$ as in \eqref{Dirac_CotY}. Then the following statements are equivalent:
\begin{itemize}
\item The curve $\mathrm{x}(t)=(t,x(t),v(t), \mathsf{p}(t), p(t)) \in \mathcal{P}=(\mathbb{R} \times T\mathcal{Q}) \times_{Y} T^{\ast}Y$  satisfies the implicit first order differential-algebraic equations
\begin{equation*}\label{implicit_DiracSys_CotY} 
\left\{
\begin{array}{l}
\displaystyle \vspace{0.2cm}\dot{x}=v, \quad  \dot{\mathsf{p}}-\frac{\partial L}{\partial t}(t,x,v) =\lambda_{r}\mathcal{B}^{r}(t,x,p),\quad \dot{p}-\frac{\partial L}{\partial x}(t,x,v)=\lambda_{r} \mathcal{A}^{r}(t,x,p),\\
\displaystyle \vspace{0.2cm}\mathcal{A}^{r}_{i}(t,x,p)\dot{x}^{i} +\mathcal{B}^{r}(t,x,p)=0,\quad \dot{t}=1,\\
\displaystyle p=\frac{\partial L}{\partial v}\,\quad \mathsf{p}= L- \frac{\partial L}{\partial v^i}v^i.
\end{array} \right.
\end{equation*} 
\item The curve $\mathrm{x}(t)=(t,x(t),v(t), \mathsf{p}(t), p(t)) \in \mathcal{P}=(\mathbb{R} \times T\mathcal{Q}) \times_{Y} T^{\ast}Y$  satisfies the time-dependent Lagrange-Dirac system
\[
\big((\dot{t}, \dot{x}, \dot{\mathsf{p}}, \dot{p}),\mathbf{d}_{D}L( t, x, v)\big)\in 
D_{\Delta_{T^*Y}}(t,x,\mathsf{p}, p).
\]
\end{itemize}
Moreover, the Lagrange-Dirac system implies the equations in \eqref{local_Dirac_Eqn_final} for the curve $x(t)\in\mathcal{Q}$.
\end{theorem}

\paragraph{Energy balance equation.} By using the second equations in \eqref{LD_equations} and \eqref{BPC}, we get the \textit{energy balance equation} along the solution curve $(x(t),v(t),p(t)) \in T\mathcal{Q} \oplus T^{\ast}\mathcal{Q}$ as
\[
\frac{d}{dt}E_{L}(t,x,v)=-\frac{\partial L}{\partial t}(t,x,v) -\lambda_{r}\mathcal{B}^{r}(t,x,p).
\]

\subsection{Hamilton-Dirac systems on the cotangent bundle}

Recall that the Lagrangian $L:\mathbb{R} \times T\mathcal{Q}\rightarrow\mathbb{R}$ is hyperregular if
\[
\mathbb{F}L: \mathbb{R} \times T\mathcal{Q} \rightarrow \mathbb{R} \times T^{\ast}\mathcal{Q}, \;\left(t, x, v\right) \mapsto \left(t, x, \frac{\partial L}{\partial v} \right)
\]
is a diffeomorphism. In this case we can define the time-dependent Hamiltonian $H:\mathbb{R} \times T^{\ast}\mathcal{Q}\rightarrow\mathbb{R}$ by
\[
H=E_{L} \circ (\mathbb{F}L)^{-1}.
\]
Then we will also introduce the \textit{covariant Hamiltonian} $\mathcal{H}:T^{\ast}Y\rightarrow \mathbb{R}$ given by
\begin{equation}\label{GenHam}
\mathcal{H}(t,x,\mathsf{p},p)=\mathsf{p}+ H(t, x, p).
\end{equation}
This definition comes from the context of Hamiltonian field theories. Indeed, in the general setting of field theories, see Remark \ref{field_theory}, a covariant Hamiltonian density is a map $\mathscr{H}: J^1Y^\star \rightarrow \Lambda^{n+1}X$ of the form $\mathscr{H}(x^\mu, y^i,\mathsf{p}, p^\mu_i)= (\mathsf{p}+ H(x^\mu, y^i,p^\mu_i))d^{n+1}x$, for some function $H(x^\mu, y^i,p^\mu_i)$. In the particular case $Y=\mathbb{R}\times \mathcal{Q}$, the covariant Hamiltonian density is thus of the form $\mathscr{H}(t,x,\mathsf{p},p)= \mathcal{H}(t,x,\mathsf{p},p)dt$ with $\mathcal{H}$ given in \eqref{GenHam}.

\paragraph{Hamilton-Dirac systems on $T^{\ast}Y$.} Given a variational constraint $C_V$, we consider the induced distribution $\Delta_{T^*Y}$ on $T^*Y$ defined in \eqref{def_D_CotY} and the Dirac structure $D_{\Delta_{T^*Y}}$ defined in \eqref{Dirac_CotY}. Consider a time-dependent Hamiltonian $H:\mathbb{R} \times T^*\mathcal{Q}\rightarrow \mathbb{R}$ and the associated covariant Hamiltonian $\mathcal{H}$.

\begin{definition}\label{def_Ham_Dir}
Given $\Delta_{T^*Y}$ and $H$ as above, the associated  {\bfi time-dependent Hamilton-Dirac dynamical system} for a curve
$\mathrm{z}(t)= (t,x(t),\mathsf{p}(t), p(t))\in T^*Y$ is
\begin{equation}\label{Ham_Dir_syst}
\big(\dot{\mathrm{z}}(t), \mathbf{d}\mathcal{H}(\mathrm{z}(t))\big)\in 
D_{\Delta_{T^*Y}}(\mathrm{z}(t)).
\end{equation}
\end{definition}

Using the expression \eqref{local_Dirac_P_new} of the Dirac structure \eqref{Dirac_CotY} and the expression of the differential of $\mathbf{d}\mathcal{H}: T^{\ast}Y \to T^{\ast}T^{\ast}Y$ locally given by
\[
\mathbf{d}\mathcal{H}( t, x, \mathsf{p}, p)=\left( t, x, \mathsf{p}, p, \frac{\partial H}{\partial t}, \frac{\partial H}{\partial x}, 1, \frac{\partial H}{\partial p} \right),
\]
we get from \eqref{Ham_Dir_syst} the equations
\begin{equation}\label{local_HamDiracSys}
\begin{split}
&\dot{x}^i=\frac{\partial H}{\partial p_i}, \;\; \dot{t}=1,\;\; \dot{\mathsf{p}}+\frac{\partial H}{\partial t}(t,x,p) =\lambda_{r}\mathcal{B}^{r}(t,x,p),\\[1mm]
& \dot{p}_i+\frac{\partial H}{\partial x^i}(t,x,p)=\lambda_{r} \mathcal{A}^{r}_i(t,x,p), \;\;\mathcal{A}_{i}^{r}(t,x,p)\frac{\partial H}{\partial p_{i}} +\mathcal{B}^{r}(t,x,p)=0.
\end{split}
\end{equation}

By recalling the construction of the Dirac structure $D_{\Delta_{T^*Y}}$ from a given variational constraint $C_V$, the Hamilton-Dirac formulation of the equations with nonlinear time-dependent constraint of thermodynamic type can be summarized as follows.

\begin{theorem}\label{theorem_HD}
Consider a variational constraint $\mathscr{C}_{V} \subset  (\mathbb{R} \times T^{\ast}\mathcal{Q}) \times_{Y} TY$ associated to $C_V$ as in \eqref{scrCV}, and define the induced Dirac structure $D_{\Delta_{T^*Y}}$ as in \eqref{Dirac_CotY}. Given a Hamiltonian $H$ on $\mathbb{R} \times T^{\ast}\mathcal{Q}$, define the covariant Hamiltonian $\mathcal{H}: T^{\ast}Y \to \mathbb{R}$ as in \eqref{GenHam}. Then the following statements are equivalent:
\begin{itemize}
\item The curve $\mathrm{z}(t)=(t,x(t), \mathsf{p}(t), p(t)) \in T^{\ast}Y$  satisfies the implicit first-order differential-algebraic equations:
\begin{equation*}\label{implicit_DiracSys_CotY} 
\left\{
\begin{array}{l}
\displaystyle \vspace{0.2cm}
\dot{x}^i=\frac{\partial H}{\partial p_i}, \;\; \dot{\mathsf{p}}+\frac{\partial H}{\partial t}(t,x,p) =\lambda_{r}\mathcal{B}^{r}(t,x,v),\;\;\dot{p}_i+\frac{\partial H}{\partial x^i}(t,x,p)=\lambda_{r} \mathcal{A}^{r}_i(t,x,p),\\
\displaystyle \mathcal{A}_{i}^{r}(t,x,p)\frac{\partial H}{\partial p_{i}} +\mathcal{B}^{r}(t,x,p)=0.
\end{array} \right.
\end{equation*} 
\item The curve $\mathrm{z}(t)=(t,x(t),\mathsf{p}(t), p(t)) \in T^{\ast}Y$  satisfies the time-dependent Hamilton-Dirac system
\[
\big(\dot{\mathrm{z}}(t), \mathbf{d}\mathcal{H}(\mathrm{z}(t))\big)\in 
D_{\Delta_{T^*Y}}(\mathrm{z}(t)).
\]
\end{itemize}
Moreover, the Hamilton-Dirac system yields equations that are equivalent to \eqref{local_Dirac_Eqn_final} when the Lagrangian is hyperregular.
\end{theorem} 

\paragraph{Energy balance equations.} One checks that the covariant Hamiltonian is conserved along the solution of the Hamilton-Dirac system \eqref{local_HamDiracSys}, 
\begin{equation}\label{conserv_covH}
\frac{d}{d t}\mathcal{H}(t,x(t),\mathsf{p}(t),p(t))=0.
\end{equation}
Note however that $\mathcal{H}$ does not represent the total energy of the system, which is given by $H$. In terms of $H$, the balance equation \eqref{conserv_covH} yields
\[
\frac{d}{dt}H(t,x,p)=- \frac{d}{dt}\mathsf{p}= \frac{\partial H}{\partial t}(t,x,p) -\lambda_{r}\mathcal{B}^{r}(t,x,p).
\]
Energy is not conserved, consistently with the fact that the equations describe the dynamics of an open system, see \S\ref{Section_4}. The quantity $\frac{d}{dt}\mathsf{p}$ is interpreted as the power flowing out of the system. The equation for $\mathsf{p}$ is independent  from the others. A natural choice of the initial condition of $\mathsf{p}$ is $\mathsf{p}(0)= -H(0)$, so that $\mathcal{H}(t,x,\mathsf{p},p)=0$ for all $t$, which is called the {\it super-Hamiltonian constraint}.

\paragraph{A first Hamilton-d'Alembert-Pontryagin principle.}
In order to develop the variational structure underlying the Hamilton-Dirac dynamical system with the time-dependent nonholonomic constraints of thermodynamic type, we begin with the Hamilton-d'Alembert-Pontryagin principle for curves $(x(t), p(t))\in T^{\ast}\mathcal{Q}$, which is a critical condition 
\begin{equation}\label{HDAP_TNS}
\delta \int_{t_{1}}^{t_{2}} \Big[\big\langle p(t) ,\dot{x}(t)\big\rangle- H(t,x(t),p(t))\Big]dt=0,
\end{equation}
for variations $\delta x$ and $\delta p$, with $\delta x$ subject to the variational constraint 
\[
\mathcal{A}_{i}^{r}(t,x,p)\delta{x}^{i} =0,\;\; r=1,...,m
\]
with $\delta{x}(t_{1})=\delta{x}(t_{2})=0$ and also subject to the nonlinear constraint
\[
\mathcal{A}^{r}_{i}(t,x,p)\dot x^i +\mathcal{B}^{r}(t,x,p)=0,\;\; r=1,...,m.
\]

The principle \eqref{HDAP_TNS} yields the equations of motion:
\begin{equation}\label{HdAP_conditions}
\dot x=\frac{\partial H}{\partial p} , \quad \dot p=- \frac{\partial H}{\partial x}+\lambda_{r}\mathcal{A}^{r}(t,x,p),\quad \mathcal{A}_{i}^{r}(t,x, p)\frac{\partial H}{\partial p_{i}} +\mathcal{B}^{r}(t,x, p)=0.
\end{equation}

We note that the principle \eqref{HDAP_TNS} does not yield all the equations associated to the Hamilton-Dirac system in \eqref{local_HamDiracSys}, namely, the equation for $\mathsf{p}$ is missing. Before formulating a principle that includes this equation, we shall first express the Hamilton-d'Alembert-Pontryagin principle \eqref{HDAP_TNS} in the intrinsic form by using the canonical one-form $\Theta_{T^*Y}=p_{i}dx^{i}+\mathsf{p}dt$ on $T^*Y$ and the covariant Hamiltonian $\mathcal{H}: T^{\ast}Y \to \mathbb{R}$. The Hamilton-d'Alembert-Pontryagin principle in \eqref{HDAP_TNS} can be intrinsically written as
\begin{equation}\label{intrinsic_HDAP}
\delta \int_{t_{1}}^{t_{2}} \Big[ \big\langle\Theta_{T^*Y}(\mathrm{z}(t)), \dot{\mathrm{z}}(t)\big\rangle-\mathcal{H}(\mathrm{z}(t))\Big]dt=0,
\end{equation}
subject to the kinematic and variational constraints
\begin{equation}\label{intrinsic_CV_CK_z}
\dot{\mathrm{z}}(t) \in \Delta_{T^*Y}(\mathrm{z}(t))\quad\text{and}\quad 
\delta{\mathrm{z}}(t) \in \Delta_{T^*Y}(\mathrm{z}(t))
\end{equation}
with the endpoint conditions $T\pi_{(T^*Y,Y)}(\delta\mathrm{z}(t_{1}))=T\pi_{(T^*Y,Y)}(\delta\mathrm{z}(t_{2}))=0 $.

As before, the stationary conditions for \eqref{intrinsic_HDAP} are given in \eqref{HdAP_conditions} and do not recover the equation for $\mathsf{p}$ in \eqref{local_HamDiracSys}.
This is due to the fact that $\delta \mathrm{z}$ is a vertical vector, namely $\delta \mathrm{z}= (0, \delta x, \delta p, \delta \mathsf{p})$.

\paragraph{The Hamilton-d'Alembert-Pontryagin principle associated to the Hamilton-Dirac system.} In order to recover all the equations  in \eqref{local_HamDiracSys} for the Hamilton-Dirac system from the variational formulation, we shall define the action functional \eqref{intrinsic_HDAP} for arbitrary curves $\mathrm{z}(\tau)$ in $T^*Y$, namely,
\begin{equation}\label{z_explicit}
\mathrm{z}(\tau)= (t(\tau), x(\tau), p(\tau), \mathsf{p}(\tau))\in T^*Y,
\end{equation}
rather than just on sections, while we still require that the critical curve is a section, i.e. $t(\tau)=\tau$. We thus get
\begin{equation}\label{intrinsic_HDAP_tau}
\delta \int_{\tau_{1}}^{\tau_{2}} \Big[ \big\langle\Theta_{T^*Y}(\mathrm{z}(\tau)),\mathrm{z}'(\tau)\big\rangle-\mathcal{H}(\mathrm{z}(\tau))\Big]d\tau=0,
\end{equation}
subject to the kinematic and variational constraints
\[
\dot{\mathrm{z}}(t) \in \Delta_{T^*Y}(\mathrm{z}(t))\quad\text{and}\quad 
\delta{\mathrm{z}}(t) \in \Delta_{T^*Y}(\mathrm{z}(t))
\]
with the endpoint conditions $T\pi_{(T^*Y,Y)}(\delta\mathrm{z}(t_{1}))=T\pi_{(T^*Y,Y)}(\delta\mathrm{z}(t_{2}))=0 $. Using that $\delta\mathrm{z}(t)$ is an arbitrary variation in $\Delta_{T^*Y}(\mathrm{z}(t))$ (i.e. not necessarily vertical), we get the conditions
\begin{equation}\label{intrinsic_HDAPEqn}
\mathbf{i}_{\dot{z}(t)}\Omega_{T^{\ast}Y}(z(t))-\mathbf{d}\mathcal{H}(z(t)) \in 
\Delta_{T^{\ast}Y}(z)^{\circ}, \qquad \dot{z}(t) \in \Delta_{T^{\ast}Y}(z(t))
\end{equation}
which are exactly the conditions given by the Hamilton-Dirac system \eqref{Ham_Dir_syst}.
\medskip

In local coordinates, the Hamilton-d'Alembert principle is given by the critical condition for arbitrary curves $z(\tau)$ as
\[
\delta \int_{\tau_{1}}^{\tau_{2}} \Big[\big\langle p ,x'\big\rangle +\mathsf{p} t'-\mathcal{H}(t ,x ,\mathsf{p} ,p )\Big]d\tau=0,
\]
subject to the constraint
\[
\mathcal{A}_{i}^{r}(t,x,p)\dot x^i +\mathcal{A}^{r}(t,x,p)=0, \;\; r=1,...,m.
\]
for variations subject to the variational constraint
\[
\mathcal{A}_{i}^{r}(t,x,p)\delta{x}^{i}+\mathcal{B}^{r}(t,x,p)\delta{t}=0, \;\; r=1,...,m.
\]

One directly computes that all the conditions of the Hamilton-Dirac system in \eqref{local_HamDiracSys} are recovered.

\section{Dirac formulations for open thermodynamic systems}\label{Section_4}

In this section we describe the Dirac system formulation for open thermodynamic systems, by using the formulation developed for time-dependent nonholonomic constraints of thermodynamic type in \S\ref{Section_3}.

\subsection{Geometric setting for simple open thermodynamics}
We consider an open thermodynamic system, as described in \S\ref{Section_Vari}. In particular, as in Fig.\ref{FluidPiston2}, the system is described by a Lagrangian $\mathsf{L}(q, \dot q, S,N)$ which depends on the mechanical variables $(q, \dot q)\in TQ$ as well as the thermodynamic variables, i.e., entropy $S$ and number of moles $N$ of the system. The system has $A$ ports denoted $a=1,...,A$ through which matter can flow in or out of the system. For simplicity, we do not consider the external heat sources, though they can be easily incorporated.

\paragraph{Lagrangian and constraints for open thermodynamic systems.} As explained in \S\ref{subsec_gen_setting}, in this situation, we  choose $\mathcal{Q}$ and $L:\mathbb{R}\times T\mathcal{Q}\rightarrow\mathbb{R}$ as
\[
\mathcal{Q}= Q\times \mathbb{R}^5\ni x=(q,S,N,\Gamma, W, \Sigma),
\]
\begin{equation}\label{Lagr_choice}
L(t,x,\dot x)=\mathsf{L}(q, \dot q, S,N) + \dot W N+  \dot{\Gamma }( S- \Sigma )
\end{equation}
and also choose the coefficients $A^r_i(t,x,\dot x)$ and $B^r_i(t,x,\dot x)$ in the constraints  such that
\begin{align*}
A^r_i(t,x,\dot x)\delta x^i&= - \frac{\partial \mathsf{L}}{\partial S} \delta \Sigma  +  \left< F^{\rm fr }, \delta q \right>   +\sum_{a=1}^A \Big[\mathcal{J}^a\delta W +\mathcal{J}_S^a\delta \Gamma\Big], \\
B^r_i(t,x,\dot x)&=- \sum_{a=1}^A \Big[\mathcal{J}^a \mu^a- \mathcal{J}_S^aT^a\Big].
\end{align*}

Now, we employ the local coordinates $(t,x,v)\in  \mathbb{R} \times T \mathcal{Q}$ with $x=(q,S,N,\Gamma, W, \Sigma)$ and $v=(v_q,v_S,v_N,v_\Gamma, v_W, v_\Sigma)\in T_x\mathcal{Q}$, the local coordinates $(t,x,\delta{t},\delta{x}) \in TY$ with $\delta x=(\delta q,\delta S,\delta N,\delta \Gamma, \delta W, \delta \Sigma)\in T_x\mathcal{Q}$ and,  the local coordinates $(t,x,\mathsf{p},p) \in T^{\ast}Y$ with $p=(p_q,p_S,p_N,p_\Gamma, p_W, p_\Sigma)\in T^*_x\mathcal{Q}$.

\subsection{Dirac formulation on the covariant Pontryagin bundle}\label{open_thermo_Pontryagin}

As before, let $\mathcal{P}=(\mathbb{R} \times T\mathcal{Q}) \times_{Y} T^{\ast}Y$ be the covariant Pontryagin bundle over $Y$, whose coordinates are given by $\mathrm{x}=(t,x,v, \mathsf{p},p) \in \mathcal{P}$. Recall that here $\mathcal{Q}=Q\times \mathbb{R}^5\ni x=(q,S, N, \Gamma, W, \Sigma)$.

From the canonical forms on $T^\ast Y$, the one-form  and presymplectic form  induced on $\mathcal{P}$ as $\Theta_\mathcal{P}=\pi_{(\mathcal{P}, T^{\ast}Y)}^{\ast}\Theta_{T^{\ast}Y}$ and $\Omega_\mathcal{P}=\pi_{(\mathcal{P}, T^{\ast}Y)}^{\ast}\Omega_{T^{\ast}Y}$, whose local expressions are respectively given by
\begin{align*}
\Theta_{\mathcal{P}}&=p_q dq+ p_{S}dS+ p_{N}dN+ p_{\Gamma}d\Gamma + p_{W}dW+ p_{\Sigma}d\Sigma+\mathsf{p}dt,\\
\Omega_{\mathcal{P}}&=dq\wedge dp_q + dS \wedge dp_{S}+ dN \wedge dp_{N}+ d\Gamma \wedge dp_{\Gamma}+ dW \wedge dp_{W}+ d\Sigma \wedge dp_{\Sigma}+dt \wedge d\mathsf{p}.
\end{align*}

\paragraph{The variational and kinematic constraints.} By using the definition of the variational constraint given in \eqref{CV}, we have
\begin{equation}\label{thermo_CV}
\begin{aligned} 
C_{V}&=\Big\{(t, x, v,  \delta{t}, \delta{x}) \in (\mathbb{R} \times T\mathcal{Q}) \times_{Y} TY \,\Big| \\
& \hspace{2cm}\frac{\partial \mathsf{L}}{\partial S}\delta \Sigma  =  \left< F^{\rm fr }, \delta q \right>  + \sum_{a=1}^A\big[\mathcal{J} ^{a}  (\delta  W-\mu^a\delta t) +     \mathcal{J} ^{a}_{S}(\delta  \Gamma-T^a\delta t)\big] \Big\},
\end{aligned}
\end{equation} 
where we note that the affine part of the constraint is now associated to $\delta t$. Following the construction of the kinematic constraint $C_K$ given in \eqref{CK_CV}, i.e.,
\[
C_{K}=\big\{(t,x,\dot t,\dot x)\in TY \mid (t,x,\dot t,\dot x) \in C_{V}(t,x,\dot x) \big\} \subset TY,
\]
we obtain the constraint
\begin{equation}\label{thermo_CK} 
C_{K}=\Big\{
(t,x,\dot{t},\dot{x}) \in TY \,\Big|\, \frac{\partial \mathsf{L}}{\partial S}\dot \Sigma  =  \left< F^{\rm fr }, \dot q \right> +  \sum_{a=1}^A\big[\mathcal{J} ^{a} ( \dot  W -\mu^a\dot t)+     \mathcal{J} ^{a}_{S}(\dot  \Gamma- T^a\dot t)\big] \Big\},
\end{equation} 
where we note that the affine part of the constraint is now associated with $\dot t$.

\paragraph{Dirac structures on $\mathcal{P}$ for open thermodynamic systems.} Recall from \eqref{Delta_P} that the variational constraint $C_V$ induces a distribution $\Delta_\mathcal{P}$ on $\mathcal{P}$. As shown in \eqref{Dirac_P}, from the distribution $\Delta_{\mathcal{P}}$ and the presymplectic form $\Omega_{\mathcal{P}}$, we can  define the induced Dirac structure $D_{\Delta_{\mathcal{P}}} \subset T\mathcal{P} \oplus T^{\ast}\mathcal{P}$ on $\mathcal{P}$. 

We now describe the Dirac structure induced by the variational constraint \eqref{thermo_CV}.
For each $\mathrm{x}=(t,x,v, \mathsf{p},p)\in \mathcal{P}$, we use the notation
\[
\mathfrak{u}_{\mathrm{x}}=( \dot{t}, \dot{x}, \dot{v}, \dot{\mathsf{p}}, \dot{p})\in T_{\mathrm{x}}\mathcal{P} \quad \textrm{and}\quad
\mathfrak{a}_{\mathrm{x}}=(\pi, \alpha, \beta, \gamma, w) \in T^{\ast}_{\mathrm{x}}\mathcal{P}, 
\]
where $\dot{v}=( \dot v_q, \dot{v}_{S},\dot{v}_{N},\dot{v}_{\Gamma} , \dot{v}_{W}, \dot{v}_{\Sigma})$, $\dot{p}=(\dot p_q, \dot{p}_{S},  \dot{p}_{N}, \dot{p}_{\Gamma}, \dot{p}_{W},\dot{p}_{\Sigma})$, $\alpha=(\alpha_q, \alpha_{S},\alpha_{N}, \alpha_{\Gamma}, \alpha_{W},\alpha_{\Sigma})$,  $\beta=(\beta_q, \beta_{S},\beta_{N}, \beta_{\Gamma}, \beta_{W},\beta_{\Sigma})$, and  $w=(w_q, w_{S},w_{N}, w_{\Gamma}, w_{W},w_{\Sigma})$.

We utilize the expression of the annihilator $C_V(t,x,v)^\circ$, which is given by all covectors $(t,x,\pi, \alpha)\in T^*Y$ such that
\begin{equation}\label{annihilator_concrete}
\begin{aligned}
&\pi= \frac{\alpha_\Sigma}{\frac{\partial \mathsf{L}}{\partial S}}(\mathcal{J}^a\mu^a+ \mathcal{J}_S^aT^a),\qquad \alpha_q+ \frac{\alpha_\Sigma}{\frac{\partial \mathsf{L}}{\partial S}}F^{\rm fr}=0,\\
&\alpha_S=0,\qquad \alpha_N=0,\qquad \alpha_\Gamma + \frac{\alpha_\Sigma}{\frac{\partial \mathsf{L}}{\partial S}}\mathcal{J}^a_S=0,\qquad \alpha_W+ \frac{\alpha_\Sigma}{\frac{\partial \mathsf{L}}{\partial S}}\mathcal{J}^a=0.
\end{aligned}
\end{equation}
Using this and \eqref{local_DiracCond_P}, the condition 
\[
\big(( \dot{t}, \dot{x}, \dot{v}, \dot{\mathsf{p}}, \dot{p}), (\pi, \alpha, \beta, \gamma, w)\big) \in D_{\Delta_{\mathcal{P}}}(t,x,v,\mathsf{p}, p)
\]
is given explicitly by
\begin{equation}\label{CondDiracSys_P} 
\left\{
\begin{array}{l}
\displaystyle \vspace{0.2cm} \dot{t}=\gamma, \; \dot q= w_q,\; \dot{S}=w_{S},\;\dot{N}=w_{N},\;\dot{\Gamma}=w_{\Gamma},\;\dot{W}=w_{W},\;\dot{\Sigma}=w_{\Sigma},\\[2mm]
\displaystyle \vspace{0.2cm} \beta_q=\beta_{S}=\beta_{N}=\beta_{\Gamma}=\beta_{W}=\beta_{\Sigma}=0,\\
\displaystyle \vspace{0.2cm}
\dot{\mathsf{p}}+\pi=\frac{1}{\frac{\partial \mathsf{L}}{\partial S}}(\dot{p}_{\Sigma}+\alpha_{\Sigma})\sum_{a=1}^A(\mathcal{J} ^{a}\mu^{a} +\mathcal{J} ^{a}_{S}T^{a} )=0,\\
\displaystyle \vspace{0.2cm} \dot p + \alpha_q+ \frac{1}{\frac{\partial \mathsf{L}}{\partial S}}(\dot{p}_{\Sigma}+\alpha_\Sigma)F^{\rm fr},\\
\displaystyle \vspace{0.2cm} \dot{p}_{S}+\alpha_{S}=0,\qquad  \dot{p}_{N}+\alpha_{N}=0,\\
\displaystyle \vspace{0.2cm} \dot{p}_{\Gamma}+\alpha_{\Gamma} + \frac{1}{\frac{\partial \mathsf{L}}{\partial S}}(\dot{p}_{\Sigma}+\alpha_{\Sigma})\sum_{a=1}^A\mathcal{J} ^{a}_S=0,\\
\displaystyle \vspace{0.2cm} \dot{p}_{W}+\alpha_{W} + \frac{1}{\frac{\partial \mathsf{L}}{\partial S}}(\dot{p}_{\Sigma}+\alpha_{\Sigma})\sum_{a=1}^A\mathcal{J} ^{a}=0,\\
\displaystyle \vspace{0.2cm}\frac{\partial \mathsf{L}}{\partial S}\dot \Sigma  =  \left< F^{\rm fr }, \dot q \right> +  \sum_{a=1}^A\big[\mathcal{J} ^{a} ( \dot  W -\mu^a\dot t)+     \mathcal{J} ^{a}_{S}(\dot  \Gamma- T^a\dot t)\big].
\end{array} \right.
\end{equation} 
Expression \eqref{CondDiracSys_P} is the local description of the Dirac structure $D_{\Delta_\mathcal{P}}\subset T\mathcal{P}\oplus T^*\mathcal{P}$.

\paragraph{Dirac system on $\mathcal{P}$ for open thermodynamic systems.} Recall from \eqref{Lagr_choice} that the Lagrangian to be considered is given by $L(t,x,\dot x)=\mathsf{L}(q, \dot q, S,N) + \dot W N+  \dot{\Gamma }( S- \Sigma )$. The covariant generalized energy, see \eqref{CovariantEnergy}, is here given by
\begin{align*}
&\mathcal{E}(t,x,v,\mathsf{p},p)=\mathsf{p}+\left<p, v \right>-L(t,x,v)\\
&=\mathsf{p} + \left<p_q, v_q \right> + p_{S}v_{S}+p_{N}v_{N}+ (p_{\Gamma}+\Sigma - S) v_{\Gamma}+ (p_W-N)v_W+ p_\Sigma v_\Sigma - \mathsf{L}(q,v_q,S,N).
\end{align*}
The differential of $\mathbf{d}\mathcal{E}$ is obtained by
\[
\mathbf{d}\mathcal{E}(t,x,v,\mathsf{p},p)=\left( -\frac{\partial L}{\partial t}, -\frac{\partial L}{\partial x}, p-\frac{\partial L}{\partial v}, 1, v \right)=(\pi, \alpha, \beta, \gamma, w),
\]
where
\[
\pi=-\frac{\partial  L}{\partial t}=0,\qquad  \alpha=-\frac{\partial L}{\partial x}=\left(-\frac{\partial\mathsf{L}}{\partial q},-v_\Gamma -\frac{\partial \mathsf{L}}{\partial S},-v_W -\frac{\partial \mathsf{L}}{\partial N}, 0, 0,v_\Gamma\right)
\]
and
\[
\beta=p-\frac{\partial L}{\partial v}=\left(p_q-\frac{\partial \mathsf{L}}{\partial v_q}, p_S, p_N, p_\Gamma+\Sigma-S, p_W-N, p_\Sigma\right),\qquad w=v=(v_q, v_S, v_N, v_\Gamma, v_W, v_\Sigma).
\]

By using this and the expression \eqref{CondDiracSys_P} of the Dirac structure, it follows that the Dirac dynamical system 
$\big(( \dot{t}, \dot{q}, \dot{v}, \dot{\mathsf{p}}, \dot{p}), \mathbf{d}\mathcal{E}(t,q,v,\mathsf{p},p)\big) \in D_{\Delta_{\mathcal{P}}}(t,q,v,\mathsf{p},p)$ is equivalent to
\begin{equation}\label{CondDiracSystem_P} 
\left\{
\begin{array}{l}
\displaystyle \vspace{0.2cm} \dot{t}=1, \; \dot q= v_q,\; \dot{S}=v_{S},\;\dot{N}=v_{N},\;\dot{\Gamma}=v_{\Gamma},\;\dot{W}=v_{W},\;\dot{\Sigma}=v_{\Sigma},\\[3mm]
\displaystyle \vspace{0.2cm} p_q-\frac{\partial \mathsf{L}}{\partial v_q}=0, \; p_S=0,\; p_N=0,\; p_\Gamma+\Sigma-S=0,\; p_W-N=0,\; p_\Sigma=0,\\[2mm]
\displaystyle \vspace{0.2cm}
\dot{\mathsf{p}}+0=\frac{1}{\frac{\partial \mathsf{L}}{\partial S}}(\dot{p}_{\Sigma}+v_\Gamma)\sum_{a=1}^A(\mathcal{J} ^{a}\mu^{a} +\mathcal{J} ^{a}_{S}T^{a} )=0,\\
\displaystyle \vspace{0.2cm} \dot p_q - \frac{\partial\mathsf{L}}{\partial q}+ \frac{1}{\frac{\partial \mathsf{L}}{\partial S}}(\dot{p}_{\Sigma}+v_\Gamma)F^{\rm fr}=0,\\
\displaystyle \vspace{0.2cm} \dot{p}_{S}-v_\Gamma -\frac{\partial \mathsf{L}}{\partial S}=0,\qquad  \dot{p}_{N}-v_W-\frac{\partial \mathsf{L}}{\partial N}=0,\\
\displaystyle \vspace{0.2cm} \dot{p}_{\Gamma}+0+ \frac{1}{\frac{\partial \mathsf{L}}{\partial S}}(\dot{p}_{\Sigma}+v_\Gamma)\sum_{a=1}^A\mathcal{J} ^{a}_S=0,\qquad \dot{p}_{W}+0+ \frac{1}{\frac{\partial \mathsf{L}}{\partial S}}(\dot{p}_{\Sigma}+v_\Gamma)\sum_{a=1}^A\mathcal{J} ^{a}=0,\\
\displaystyle \vspace{0.2cm}\frac{\partial \mathsf{L}}{\partial S}\dot \Sigma  =  \left< F^{\rm fr }, \dot q \right> +  \sum_{a=1}^A\big[\mathcal{J} ^{a} ( \dot  W -\mu^a\dot t)+     \mathcal{J} ^{a}_{S}(\dot  \Gamma- T^a\dot t)\big].
\end{array} \right.
\end{equation} 
Since $p_S=0$, we have $v_\Gamma=-\frac{\partial \mathsf{L}}{\partial S}$ from the equation in the fifth line. From this and $p_\Sigma=0$, we obtain $\frac{1}{\frac{\partial \mathsf{L}}{\partial S}}(\dot{p}_{\Sigma}+v_\Gamma)=-1$. 

Making rearrangements, we  get the following evolution equations:
\begin{equation}\label{intermediate_system} 
\left\{
\begin{array}{l}
\displaystyle \vspace{0.2cm} p_q=\frac{\partial \mathsf{L}}{\partial \dot q},\quad p_\Gamma= S-\Sigma,\quad p_W=N,\\
\displaystyle \vspace{0.2cm} \dot{\mathsf{p}}=-\sum_{a=1}^A(\mathcal{J} ^{a}\mu^{a} +\mathcal{J} ^{a}_{S}T^{a}),\quad \dot p_q=\frac{\partial L}{\partial q}+ F^{\rm fr},\quad \dot p_\Gamma= \sum_{a=1}^A\mathcal{J}^a_S,\quad \dot p_W= \sum_{a=1}^A\mathcal{J}^a,\\
\displaystyle \vspace{0.2cm} \dot\Gamma= -\frac{\partial \mathsf{L}}{\partial S},\quad \dot W= -\frac{\partial \mathsf{L}}{\partial N},\\
\displaystyle \vspace{0.2cm}\frac{\partial \mathsf{L}}{\partial S}\dot \Sigma  =  \left< F^{\rm fr }, \dot q \right> +  \sum_{a=1}^A\big[\mathcal{J} ^{a} ( \dot  W -\mu^a)+     \mathcal{J} ^{a}_{S}(\dot  \Gamma- T^a)\big].
\end{array} \right.
\end{equation} 
This formula is quite useful for making physical interpretations for the variables, as will be shown below.
By further rearrangements, we finally get the required evolution equations:
\begin{equation}\label{open_system_final} 
\left\{
\begin{array}{l}
\displaystyle\vspace{0.2cm}\frac{d}{dt}\frac{\partial \mathsf{L}}{\partial \dot q}- \frac{\partial \mathsf{L}}{\partial q}= F^{\rm fr} ,\quad\quad\quad  \frac{d}{dt} N= \sum_{a=1}^A \mathcal{J}^a,\\
\displaystyle\vspace{0.2cm}\frac{\partial \mathsf{L}}{\partial S}\Big(\dot S -\sum_{a=1}^A \mathcal{J}_S^a\Big)=  \left<F^{\rm fr}, \dot q \right> -\sum_{a=1}^A\left[\mathcal{J}^a\Big(\frac{\partial \mathsf{L}}{\partial N}+\mu^a\Big)+\mathcal{J}^a_S\Big(\frac{\partial \mathsf{L}}{\partial S}+ T^a\Big)\right].
\end{array} \right.
\end{equation} 
Notice that this recovers the equations \eqref{open_system} for the open system for the case in which there exist no external heat sources.

\paragraph{Interpretation of the thermodynamic variables.}
Note that the two equations in the third line of \eqref{intermediate_system} attribute to the variables $\Gamma$ and $W$ the meaning of thermodynamic displacements associated to the process of heat and matter transport. The conjugate momenta $p_W=N$ associated to $W$ is interpreted as the number of moles in the system, whose rate of change is indeed given by
\[
\dot p_W=\sum_{a=1}^A\mathcal{J}^a,
\]
from the fourth equation in the second line of \eqref{intermediate_system}.
The conjugate momenta associated to $\Gamma$ is given by $p_\Gamma= S-\Sigma$ and corresponds to the part of the entropy of the system that is due to the exchange of entropy with exterior. From the third equation in the second line of \eqref{intermediate_system}, its rate of change is indeed
\[
\dot p_\Gamma= \dot S-\dot\Sigma= \sum_{a=1}^A\mathcal{J}^a_S.
\]
Thus, the rate of the total entropy change of the system can be written as
\begin{equation}\label{total_entrop_prod}
\dot S=\dot \Sigma + \dot p_\Gamma= I +  \sum_{a=1}^A\mathcal{J}^a_S,
\end{equation}
where the internal entropy production $\dot \Sigma =I$ is positive by the second law of thermodynamics, while $\dot p_\Gamma= \sum_{a=1}^A\mathcal{J}^a_S$ is the rate of entropy flowing into the system and has an arbitrary sign.
Equation \eqref{total_entrop_prod} is often denoted in the form
\[
dS= d_iS+ d_eS
\]
in physics textbooks (see, for instance, \cite{deGrootMazur1969}), where $dS$ denotes the infinitesimal change of the total entropy,  $d_iS$ the entropy produced inside the system and $d_eS$ the entropy supplied to the system by its surroundings. In our formulation, it reads as $d_iS=\dot\Sigma dt$ and $d_eS= \dot p_\Gamma dt$.

Finally, the momentum $\mathsf{p}$ represents the part of the energy associated to the interaction of the system with the exterior through its ports. In fact, its rate of change is
\[
\frac{d}{dt}\mathsf{p}= - \sum_{a=1}^A (\mathcal{J}^a\mu^a + \mathcal{J}_S^a T^a)=-\frac{d}{dt}E= - P_M^{\rm ext},
\]
where $E$ is the total energy of the system, defined as
\[
E(q, \dot q, S, N)= \frac{\partial \mathsf{L}}{\partial \dot q^i}\dot q^i- \mathsf{L}(q,\dot q, S, N).
\]
We note that this energy coincides with the energy defined from the Lagrangian $L(t,x,\dot x)=\mathsf{L}(q, \dot q, S,N) + \dot W N+  \dot{\Gamma }( S- \Sigma )$ via the formula $E(t,x,\dot x)= \frac{\partial L}{\partial \dot x^i}\dot x^i- L(t,x,\dot x)$.

\subsection{Lagrange-Dirac formulation on the cotangent bundle}

We now quickly present the Lagrange-Dirac formulation on $T^*Y$ for the thermodynamic of open systems for the Lagrangian
$L$ on $\mathbb{R}\times T\mathcal{Q}$ given in \eqref{Lagr_choice}.

\paragraph{The variational and kinematic constraints.} 
The first step is the construction of the variational constraint $\mathscr{C}_V$ from $C_V$ by following \eqref{scrCV}.
We assume that the constraint $C_V$ in \eqref{thermo_CV} depends on $v$ only through its first component $v_q$. This hypothesis means that the quantities $F^{\rm fr}$, $\mathcal{J}^a$, $\mathcal{J}^a_S$ do not depend on $v_{S}=\dot S$, $v_N=\dot{N}$, $v_\Gamma=\dot{\Gamma}$, $v_{W}=\dot W$, and $v_{\Sigma}=\dot{\Sigma}$. This assumption is verified for realistic examples. Dependence on all the other variables $(q,v_q,S,N,W,\Gamma, \Sigma)$ is however allowed, although only dependence on $(q,v_q,S,N)$ is assumed in practice.

The definition of the variational constraint $\mathscr{C}_V$ from $C_V$ thus only needs that the Lagrangian $\mathsf{L}$ is nondegenerate with respect to the mechanical variable $v_q$. In this case, we can define the friction forces $\mathcal{F}^{\rm fr}(q,p_q,S,N):= F^{\rm fr}(q,v_q,S,N)$, where $v_q$ is such that $\frac{\partial \mathsf{L}}{\partial v_q}(q,v_q, S, N)=p_q$. Following \eqref{scrCV}, the variational constraint $\mathscr{C}_V$ is thus
\begin{equation}\label{thermo_CV_scr}
\begin{aligned} 
\mathscr{C}_{V}&=\Big\{(t, x, p,  \delta{t}, \delta{x}) \in (\mathbb{R} \times T^*\mathcal{Q}) \times_{Y} TY \,\Big| \\
& \hspace{2cm}\frac{\partial \mathsf{L}}{\partial S}\delta \Sigma  =  \mathcal{F}^{\rm fr}\!\cdot\! \delta q  + \sum_{a=1}^A\big[\mathcal{J} ^{a}  (\delta  W-\mu^a\delta t) +     \mathcal{J} ^{a}_{S}(\delta  \Gamma-T^a\delta t)\big] \Big\},
\end{aligned}
\end{equation} 
where $\mathcal{F}^{\rm fr}$, $\mathcal{J} ^{a}_{S}$, and $\mathcal{J} ^{a}_{S}$ are all expressed in terms of the variables $(q,p_q,S,N)$.

\paragraph{Dirac structures on $T^*Y$ for open thermodynamic systems.} Recall from \eqref{def_D_CotY} that the variational constraint $\mathscr{C}_V$ induces a distribution $\Delta_{T^*Y}$ on $T^*Y$. As shown in \eqref{Dirac_CotY}, from the distribution $\Delta_{T^*Y}$ and the presymplectic form $\Omega_{T^*Y}$, we can  define the induced Dirac structure $D_{\Delta_{T^*Y}} \subset TT^*Y \oplus T^{\ast}T^*Y$ on $T^*Y$. 

We now describe the Dirac structure induced by the variational constraint \eqref{thermo_CV_scr}.
For each $\mathrm{z}=(t,x,\mathsf{p},p)\in T^*Y$, we use the notation
\[
\mathfrak{u}_{\mathrm{z}}=( \dot{t}, \dot{x}, \dot{\mathsf{p}}, \dot{p})\in T_{\mathrm{z}}T^*Y \quad \textrm{and}\quad
\mathfrak{a}_{\mathrm{z}}=(\pi, \alpha, \gamma, w) \in T^{\ast}_{\mathrm{z}}T^*Y, 
\]
where $\dot{p}=(\dot p_q, \dot{p}_{S},  \dot{p}_{N}, \dot{p}_{\Gamma}, \dot{p}_{W},\dot{p}_{\Sigma})$, $\alpha=(\alpha_q, \alpha_{S},\alpha_{N}, \alpha_{\Gamma}, \alpha_{W},\alpha_{\Sigma})$, and  $w=(w_q, w_{S},w_{N}, w_{\Gamma},$ $ w_{W},w_{\Sigma})$.

We will use the expression of the annihilator $\mathscr{C}_V(t,x,p)^\circ$ which is given by the same conditions as \eqref{annihilator_concrete}, with the only change that $\mathcal{F}^{\rm fr}$, $\mathcal{J} ^{a}_{S}$, and $\mathcal{J} ^{a}_{S}$ are all expressed in terms of the variables $(q,p_q,S,N)$ rather than $(q,v_q, S, N)$.

Using this and \eqref{DiracCond_CotY}, the condition 
\[
\Big(( \dot{t}, \dot{x},  \dot{\mathsf{p}}, \dot{p}), (\pi, \alpha, \gamma, w)\Big) \in D_{\Delta_{T^*Y}}(t,x,\mathsf{p}, p)
\]
is given explicitly by the same expression as in 
\eqref{CondDiracSys_P} with all the equations in the second line removed, thereby giving the local description of the Dirac structure $D_{T^*Y}\subset TT^*Y\oplus T^*T^*Y$.

\paragraph{Lagrange-Dirac system on $T^*Y$ for open thermodynamic systems.} We shall not describe in details the Lagrange-Dirac system, as the computations are similar to those made in the previous case in \S\ref{open_thermo_Pontryagin}.
We shall just describe the Dirac differential of $L$, namely, $\mathbf{d}_{D}L : \mathbb{R} \times T\mathcal{Q} \to T^\ast T^\ast Y$, which is given by 
\[
\mathbf{d}_DL(t,x,v)=\left( t,x, -E_L, \frac{\partial L}{\partial v},-\frac{\partial L}{\partial t}, -\frac{\partial L}{\partial x}, 1, v \right)=(t,x,\mathsf{p},p, \pi, \alpha, \gamma, w),
\]
where
\[
\mathsf{p}= - E_L= - \left(\frac{\partial \mathsf{L}}{\partial v_q^i}v_q^i- \mathsf{L}\right),\qquad p=\frac{\partial L}{\partial v}=\left(\frac{\partial \mathsf{L}}{\partial v_q}, 0, 0, -\Sigma+S, N, 0\right),
\]
\[
\pi=-\frac{\partial  L}{\partial t}=0,\qquad  \alpha=-\frac{\partial L}{\partial x}=\left(-\frac{\partial\mathsf{L}}{\partial q},-v_\Gamma -\frac{\partial \mathsf{L}}{\partial S},-v_W -\frac{\partial \mathsf{L}}{\partial N}, 0, 0,v_\Gamma\right).
\]
Thus, it can be easily checked that the Lagrange-Dirac system 
$$
\left((\dot{t}, \dot x(t), \dot{\mathsf{p}}(t), \dot p(t)),\mathbf{d}_DL( t, x(t), v(t))\right)\in D_{\Delta_{T^*Y}}(t,x(t),\mathsf{p}(t), p(t))
$$
does yield the system of equations \eqref{open_system_final} for the open system.

\paragraph{Remark.} Concerning the Hamiltonian setting for open systems, it should be noted that one cannot make the Legendre transformation for the Lagrangian $L(t,x,\dot x)$ on $\mathbb{R} \times T\mathcal{Q}$ to get a Hamiltonian on $T^\ast(\mathbb{R} \times \mathcal{Q})$, since the Lagrangian $L(t,x,\dot x)=\mathsf{L}(q, \dot q, S,N) + \dot W N+  \dot{\Gamma }( S- \Sigma )$
is degenerate. However, in this case, we may apply Dirac's theory of constraints in the context of Dirac structures (see \cite{YoMa2007}), which may lead to a generalized Hamilton-Dirac system for open thermodynamics. This issue will be considered in a future work.

\section{Conclusions}

In this paper, we have shown that Dirac structures can be defined for open thermodynamic systems, specifically by focusing on simple systems. We have presented both the Dirac dynamical system formulation as well as the associated variational structures.
 In particular, we have shown that the underlying geometric structure is given by a time-dependent Dirac structure on the covariant Pontryagin bundle $\mathcal{P}= (\mathbb{R} \times T\mathcal{Q}) \times_{Y} T^\ast Y$ over the extended thermodynamic configuration manifold $Y=\mathbb{R} \times \mathcal{Q}$, where $\mathbb{R}$ denotes the space of time and $\mathcal{Q}$ the thermodynamic configuration space. This Dirac structure is induced from a variational constraint $C_V \subset (\mathbb{R}\times T\mathcal{Q})\times_Y T Y$, which gives rise to a distribution on $\mathcal{P}$. Associated with the time-dependent Dirac structure on $\mathcal{P}$, we have constructed a Dirac dynamical system by using the generalized covariant energy $\mathcal{E}$ on $\mathcal{P}$. This system produces the evolution equations for simple open thermodynamic systems. The kinematic constraint $C_K \subset TY$ is automatically deduced from the Dirac thermodynamic system because of the specific relation between $C_K$ and $C_V$. Constraints satisfying this relation are called constraints of thermodynamic type. We have shown that all the conditions given by the Dirac thermodynamic system can be obtained by the Lagrange-d'Alembert-Pontryagin principle for curves in the covariant Pontryagin bundle. Besides the Dirac formulation on the covariant Pontryagin bundle $\mathcal{P}$, we have also presented two other Dirac formulations based on the induced Dirac structure on the cotangent bundle $T^\ast Y$ over the extended configuration manifold $Y$. These are the Lagrange-Dirac and Hamilton-Dirac system, where the latter can be developed only when the Lagrangian is hyperregular.

Finally, we have illustrated our theory by the example of open thermodynamics. We have also given a physical interpretation of the thermodynamic variables involved in the Dirac system formulation, in which the well-known relation associated to the infinitesimal change of the total entropy given by $dS= d_iS+ d_eS$ can be systematically clarified. 
\medskip

As future works, we raise the following topics:
\begin{itemize}
\item Construction of Dirac structures for {\it non-simple systems}, such as those including several entropies in the system;
\item Extension to the case of continuous (infinite dimensional) nonequilibrium thermodynamic systems following \cite{GBYo2017b} and \cite{GB2019};
\item Dirac structures induced from nonholonomic kinematic constraints in addition to the nonlinear constraints of thermodynamic type;
\item Development of the Hamiltonian setting for open thermodynamic systems.
\end{itemize}

\paragraph{Acknowledgements.} H.Y. is partially supported by JSPS Grant-in-Aid for Scientific Research (17H01097), the MEXT Top Global University Project and Waseda University (SR 2019C-176,SR 2019Q-020), Interdisciplinary institute for thermal energy conversion engineering and mathematics); F.G.B. is partially supported by the ANR project GEOMFLUID, ANR-14-CE23-0002-01.


\begin{thebibliography}{xx}


\bibitem[Bloch(2003)]{Bl2003}
Bloch, A.~M. [2003], {\it Nonholonomic Mechanics and Control}, volume~24 of {\it Interdisciplinary Applied Mathematics}, Springer-Verlag, New York. With the collaboration of J. Baillieul, P. Crouch and J. Marsden, and with scientific input from P. S. Krishnaprasad, R. M. Murray and D. Zenkov.

\bibitem[Bloch and Crouch(1997)]{BlCr1997}
Bloch, A.~M. and P.~E. Crouch  [1997], Representations of {D}irac structures on
  vector spaces and nonlinear {L}--{C} circuits. In: {\em Differential Geometry
  and Control (Boulder, CO, 1997)}. Vol.~64. pp.~103--117.
\newblock Amer. Math. Soc. Providence, RI.


\bibitem[Carath\'eodory(1909)]{Ca1909}
Carath\'eodory, C. [1909], Untersuchungen \"uber die Grundlagen der Thermodynamik, \textit{Math. Ann.}, \textbf{67}, 355--386.


\bibitem[Cendra and Grillo(2007)]{CeGr2007}
Cendra, H. and S.~D. Grillo [2007], Lagrangian systems with higher order constraints, \textit{J. Math. Phys.} \textbf{48} 052904. 

\bibitem[Cendra, Ibort, de Le\'on, and Mart\'in de Diego(2004)]{CeIbdLdD2004}
Cendra, H., A. Ibort, M. de Le\'on, and D. de Diego [2004], A generalization of Chetaev's principle for a class of higher order nonholonomic constraints, \textit{J. Math. Phys.} \textbf{45}, 2785.



\bibitem[Courant(1990)]{Cour1990}
Courant, T.~J. [1990], Dirac manifolds, {\em Trans. Amer. Math. Soc.}, \textbf{319}, 631--661.

\bibitem[Courant and Weinstein(1988)]{CoWe1988}
Courant, T. and A.~Weinstein [1988], Beyond {P}oisson structures. In: {\em Action hamiltoniennes de groupes. Troisi\`eme th\'eor\`eme de Lie (Lyon, 1986)}. Vol.~27. pp.~39--49.
\newblock Hermann. Paris.

\bibitem[de Groot and Mazur(1969)]{deGrootMazur1969}
de Groot, S.~R and P. Mazur [1969], {\it Nonequilibrium Thermodynamics}, North-Holland.


\bibitem[Dorfman(1993)]{Dorfman1993}
Dorfman, I. [1993], {\em Dirac Structures and Integrability of Nonlinear Evolution Equations}.
\newblock In Nonlinear Science Theory and Applications. John Wiley \& Sons Ltd., Chichester.

\bibitem[Eberard et al.(2007)]{EbMaVa2007}
Eberard, D., B. M. Maschke, and A. J. van der Schaft [2007], An extension of Hamiltonian systems to the thermodynamic phase space: Towards a geometry of nonreversible processes, {\it Reports on Mathematical Physics} {\bf 60}(2), 175--198.

\bibitem[Ferrari and Gruber(2010)]{FeGr2010}
Ferrari, C. and C. Gruber [2010], Friction force: from mechanics to thermodynamics, \textit{Europ. J. Phys.} \textbf{31}(5), 1159--1175.


\bibitem[Fuchs(2010)]{Fu2010}
Fuchs, H.~U. [2010], \textit{The Dynamics of Heat: A Unified Approach to Thermodynamics and Heat Transfer}, Graduate Texts in Physics, Springer Science+Buisness Media. 


\bibitem[Gay-Balmaz(2019)]{GB2019}
Gay-Balmaz, F. [2019], A variational derivation of the nonequilibrium thermodynamics of a moist atmosphere with rain
process and its pseudoincompressible approximation, \textit{Geophys. Astrophys. Fluid Dyn.}, to appear. doi:10.1080/03091929.2019.1570505


\bibitem[Gay-Balmaz and Yoshimura(2017a)]{GBYo2017a}
Gay-Balmaz, F. and H. Yoshimura [2017a], A Lagrangian variational formulation for nonequilibrium thermodynamics. Part I: discrete systems, \textit{J. Geom. Phys.}\textbf{111}, 169--193.

\bibitem[Gay-Balmaz and Yoshimura(2017b)]{GBYo2017b}
Gay-Balmaz, F. and H. Yoshimura [2017b], A Lagrangian variational formulation for nonequilibrium thermodynamics. Part II: continuum systems, \textit{J. Geom. Phys.} \textbf{111}, 194--212.

\bibitem[Gay-Balmaz and Yoshimura(2018a)]{GBYo2018a}
Gay-Balmaz, F. and H. Yoshimura [2018a], A variational formulation of nonequilibrium thermodynamics for discrete open systems with mass and heat transfer, \textit{Entropy}, {\bf 20}(3), 163; doi: 10.3390/e20030163, 1--26.


\bibitem[Gay-Balmaz and Yoshimura(2018b)]{GBYo2018b}
Gay-Balmaz, F. and H. Yoshimura [2018b], Variational discretization for the nonequilibrium thermodynamics of simple systems, \textit{Nonlinearity}, in press.

\bibitem[Gay-Balmaz and Yoshimura(2018c)]{GBYo2018c}
Gay-Balmaz, F. and H. Yoshimura [2018c], Dirac structures in nonequilibrium thermodynamics, \textit{J. Math. Phys.} {\bf 59}, 012701-29.

\bibitem[Gay-Balmaz and Yoshimura(2019)]{GBYo2019}
Gay-Balmaz, F. and H. Yoshimura [2019], From Lagrangian mechanics to nonequilibrium thermodynamics: A variational perspective, \textit{Entropy} \textbf{21}(1), 8; doi: 10.3390/e21010008, 1--39.



\bibitem[Gibbs(1873a)]{Gibbs1873a}
Gibbs, J.~W. [1873a], Graphical methods in the thermodynamics of fluids, \textit{Trans. Conn. Acad.} \textbf{2}, 309--342.

\bibitem[Gibbs(1873b)]{Gibbs1873b}
Gibbs, J.~W. [1873b], A method of geometrical representation of the thermodynamic properties of substances by means of surfaces, \textit{Trans. Conn. Acad.} \textbf{2}, 382--404.


\bibitem[Gibbs(1902)]{Gibbs1902}
Gibbs, J.~W. [1902], {\it Collected Works}, Scribner, New York.


\bibitem[Glansdorff and Prigogine(1971)]{GlPr1971}
Glansdorff, P. and I. Prigogine [1971], {\it Thermodynamic Theory of Structure, Stability, and Fluctuations}, Wiley-Interscience.



\bibitem[Gotay, Isenberg, Marsden and Montgomery(1997)]{GIMM1997}
Gotay, M., J. Isenberg, J. E. Marsden and R. Montgomery [1997], Momentum maps and classical relativistic fields, Part I: Covariant field theory (66 pages). www.arxiv.org: [2004] physics/9801019.




\bibitem[Gruber(1997)]{Gr1997}
Gruber, C. [1997], \textit{Thermodynamique et M\'ecanique Statistique}, Institut de physique th\'{e}orique, EPFL. 

\bibitem[Gruber(1999)]{Gr1999}
Gruber, C. [1999], Thermodynamics of systems with internal adiabatic constraints: time evolution of the adiabatic piston, \textit{Eur. J. Phys.} \textbf{20}, 259--266. 

\bibitem[Gruber and Frachebourg(1999)]{GrFr1999}
Gruber, C. and L. Frachebourg [1999], On the adiabatic properties of a stochastic adiabatic wall: Evolution, stationary
non-equilibrium, and equilibrium states, \textit{Physica} A, \textbf{272}, 392--428. 

\bibitem[Gruber and Brechet(2011)]{GrBr2011}
Gruber, C. and S.~D. Brechet [2011], Lagrange equation coupled to a thermal equation: mechanics as a consequence of thermodynamics, \textit{Entropy} \textbf{13}, 367--378. 

\bibitem[Gyarmati(1970)]{Gyarmati1970}
Gyarmati, I. [1970], \textit{Nonequilibrium Thermodynamics: Field Theory and
Variational Principles}, Springer-Verlag, New York.

\bibitem[Hermann(1973)]{He1973}
Hermann, R. [1973], \textit{Geometry, physics and systems}, Dekker, New York.


\bibitem[Kedem and Katchalsky(1963a)]{KeKa1963a}
Kedem O. and A. Katchalsky [1963], Permeability of composite membranes. Part 1. Electric current, volume flow and flow of solute through membranes, \textit{Transactions of the Faraday Society}, \textbf{59}, 1918--1930.

\bibitem[Kedem and Katchalsky(1963b)]{KeKa1963b}
Kedem O. and A. Katchalsky [1963], Permeability of composite membranes. Part 2. Parallel elements, \textit{Transactions of the Faraday Society} \textbf{59}, 1931--1940.

\bibitem[Kedem and Katchalsky(1963c)]{KeKa1963c}
Kedem O. and A. Katchalsky [1963], Permeability of composite membranes. Part 3. Series array of elements, \textit{Transactions of the Faraday Society} \textbf{59}, 1941--1953.

\bibitem[Klein and Nellis(2011)]{KlNe2011}
Klein, S., and G. Nellis [2011], \textit{Thermodynamics}, Cambridge University Press.

\bibitem[Kosmann-Schwarzbach(2013)]{KS2013}
Kosmann-Schwarzbach, Y. [2013], Courant Algebroids. A Short History, \textit{SIGMA} \textbf{9}, 014.


\bibitem[Kondepudi and Prigogine(1998)]{KoPr1998}
Kondepudi, D. and I. Prigogine [1998], \textit{Modern Thermodynamics}, John Wiley \& Sons.



\bibitem[Lanczos(1970)]{La1970}
Lanczos, C. [1970], {\it Variational principles of mechanics}, fourth edition, Dover.


\bibitem[Lavenda(1978)]{Lavenda1978}
Lavenda, B.~H. [1978], \textit{Thermodynamics of Irreversible Processes}, Macmillan, London.

\bibitem[Machlup  and Onsager(1953)]{MaOn1953}
Onsager, L. and S. Machlup [1953], Fluctuations and irreversible processes II. Systems with kinetic energy. \textit{Phys. Rev.} \textbf{91}, 1512--1515.

\bibitem[Marle(1998)]{Ma1998}
Marle, C.-M. [1998], Various approaches to conservative and nonconservative non-holonomic systems, \textit{Rep. Math. Phys.} \textbf{42}, 1/2, 211--229.

\bibitem[Mrugala(1978)]{Mr1978}
Mrugala, R. [1978], Geometrical formulation of equilibrium phenomenological thermodynamics, \textit{Rep. Math. Phys.} \textbf{14}, 419--427.

\bibitem[Mrugala(1980)]{Mr1980}
Mrugala, R. [1980], A new representation of Thermodynamic Phase Space, \textit{Bull. Polish Acad. Sci.} \textbf{28}, 13--18.

\bibitem[Mrugala et al.(1991)]{MrNuScSa1991}
Mrugala, R., J. D. Nulton, J. C. Schon, and P. Salamon [1991], Contact structure in thermodynamic theory, \textit{Rep. Math. Phys.} \textbf{29}, 109--121.
 
\bibitem[Onsager(1931)]{Onsager1931}
Onsager, L. [1931], Reciprocal relations in irreversible processes I, \textit{Phys. Rev.} \textbf{37}, 405--426; Reciprocal relations in irreversible processes II, \textit{Phys. Rev.} \textbf{38}, 2265--2279.  

\bibitem[Onsager and Machlup(1953)]{OnMa1953}
Onsager, L. and S. Machlup [1953], Fluctuations and irreversible processes, \textit{Phys. Rev.} \textbf{91}, 1505--1512.





\bibitem[Oster, Perelson, Katchalsky(1973)]{OsPeKa1973}
Oster, G.~F., A.~S. Perelson, A. Katchalsky [1973], Network thermodynamics: dynamic modelling of biophysical systems, \textit{Quarterly Reviews of Biophysics} \textbf{6}(1), 1--134.

\bibitem[Prigogine(1947)]{Prigogine1947}
Prigogine [1947], Etude thermodynamique des ph\'enom\`enes irr\'eversibles, Thesis, Paris: Dunod and Li\`ege: Desoer.



\bibitem[Sandler(2006)]{Sa2006}
Sandler, S.~I. [2006], \textit{Chemical, Biochemical, and Engineering Thermodynamics}, John Wiley \& Sons.

\bibitem[Stueckelberg  and  Scheurer(1974)]{StSc1974}
Stueckelberg, E. C. G. and P. B. Scheurer [1974], \textit{Thermocin\'etique ph\'enom\'enologique
galil\'eenne}, Birkh\"auser, 1974

\bibitem[Tulczyjew(1977)]{Tu1977}
Tulczyjew, W.~M. [1977], The {L}egendre transformation. {\rm Ann. Inst. H. Poincar\'e} {\bf 27}(1),~101--114.

\bibitem[Truesdell(1969)]{Truesdell1969}
Truesdell, C. [1969], \textit{Rational Thermodynamics}, McGraw-Hill, New York.

\bibitem[van der Schaft and  Maschke(1995a)]{vdSMa1995a}
van der Schaft, A. J. and B. M. Maschke [1995a], The Hamiltonian formulation of energy conserving physical systems with external ports, \textit{Archiv f\"ur Elektronik und \"Ubertragungstechnik} \textbf{49}, 362--371.

\bibitem[van der Schaft and  Maschke(1995b)]{vdSMa1995b}
van der Schaft, A. J. and B. M. Maschke [1995b], Mathematical modelling of constrained Hamiltonian systems, in \textit{Proc. IFAC Symp. NOLCOS, Tahoe City, CA, International
Federation of Automatic Control}, 678--683.

\bibitem[Vankerschaver, Yoshimura, and Leok(2012)]{VaYoLe2012}
Vankerschaver, J. , H. Yoshimura, and M. Leok [2012], The Hamilton-Pontryagin principle and multi-Dirac structures for classical field theories, \textit{J. Math. Phys.} \textbf{53}, 072903-1--072903-25.

\bibitem[Woods(1975)]{Woods1975}
Woods, L.~C. [1975], \textit{The Thermodynamics of Fluid Systems}, Clarendon Press Oxford 1975.


\bibitem[Yoshimura and Marsden(2006a)]{YoMa2006a}
Yoshimura, H. and J.~E. Marsden [2006a], Dirac structures in Lagrangian mechanics. Part I: Implicit Lagrangian systems, {\rm J. Geom. and Phys.} \textbf{57}, 133--156.
  
\bibitem[Yoshimura and Marsden(2006b)]{YoMa2006b}
Yoshimura, H. and J.~E. Marsden [2006b], Dirac structures in Lagrangian mechanics. Part II: Variational structures, {\rm J. Geom. and Phys.} \textbf{57}, 209--250.

\bibitem[Yoshimura and Marsden(2007)]{YoMa2007}
Yoshimura, H. and J.~E. Marsden [2007], Dirac structures and the Legendre transformation for implicit Lagrangian and Hamiltonian systems, {\em Lagrangian and Hamiltonian Methods in Nonlinear Control 2006}, volume 366 of {\em Lecture Notes in Control and Information Science Series}, pages 233--247, Springer-Verlag.


\bibitem[Ziegler(1968)]{Ziegler1968}
Ziegler, H. [1968],  A possible generalization of Onsager's theory, in H. Barkus and L.I. Sedov (eds.), \textit{Irreversible Aspects of Continuum Mechanics}, Springer, New York.


\end{thebibliography}
\end{document}